\documentclass[useAMS,usenatbib]{mnras}
\usepackage{graphicx}
\usepackage{natbib}
\usepackage{amssymb}
\usepackage{setspace}
\usepackage{epsfig,color}
\usepackage{multicol}
\usepackage{ifpdf}
\usepackage{hyperref}
              
\newcommand{\sqc}{~cm$^{-2}$}		
\newcommand{\gsqc}{~g\,cm$^{-2}$}		
\newcommand{\cc}{~cm$^{-3}$}		
\newcommand{\gcc}{~g\,cm$^{-3}$}	
\newcommand{\das}{$.\!\!^{\prime\prime}$} 

\newcommand{\rhopdf}{$\rho$-pdf}  
\newcommand{\Npdf}{$N$-pdf}  
\newcommand{\bPLFIT}{{\sc bPlfit}}  
                                                  
\newcommand{\Hi}{H\,{\sc i}}
\newcommand{\Hii}{H\,{\sc ii}~}
\newcommand{\Htwo}{H$_2$}

\title[Power-law parts of pdfs in SF clouds]{On the extraction of power-law parts of the probability density functions in star-forming clouds}

\author[Veltchev et al.]
{	
\parbox{\textwidth}{Todor V.~Veltchev$^{1,2\,\star}$, Philipp Girichidis$^3$, Sava Donkov$^4$, Nicola Schneider$^5$, Orlin Stanchev$^1$, Lyubov Marinkova$^1$, Daniel Seifried$^5$ and  Ralf S. Klessen$^2$}\vspace{0.4cm} \\
\parbox{\textwidth}{
  $^1$University of Sofia, Faculty of Physics, 5 James Bourchier Blvd., 1164 Sofia, Bulgaria\\
  $^2$Universit\"at Heidelberg, Zentrum f\"ur Astronomie, Institut f\"ur Theoretische Astrophysik, Albert-Ueberle-Str. 2, 69120 Heidelberg, Germany\\
  $^3$Leibniz-Institut f\"ur Astrophysik Potsdam (AIP), An der Sternwarte 16, 14482 Potsdam, Germany \\
  $^4$Department of Applied Physics, Technical University, 8 Kliment Ohridski Blvd., 1000 Sofia, Bulgaria \\ 
  $^5$I. Physik. Institut, University of Cologne, Z\"{u}lpicher Str. 77, 50937 Cologne, Germany \\
  $^6$OASU/LAB, Universit\'{e} de Bordeaux, 33615 Pessac, France}
}

\date{Submitted 2019 March 18}

\ifpdf
\pagerange{\pageref{firstpage}--\pageref{lastpage}} \pubyear{2018}
\begin{document}
\label{firstpage}
\maketitle

\begin{abstract}
We present a new approach to extract the power-law part of a density/column-density probability density function (\rhopdf/\Npdf) in star-forming clouds. It is based on the mathematical method \bPLFIT{} of Virkar \& Clauset (2014) and assesses the power-law part of an arbitrary distribution, without any assumptions about the other part of this distribution. The slope and deviation point are derived as averaged values as the number of bins is varied. Neither parameter is sensitive to spikes and other local features of the tail. This adapted \bPLFIT{} method is applied to two different sets of data from numerical simulations of star-forming clouds at scales 0.5 and 500 pc and displays \rhopdf{} and \Npdf{} evolution in agreement with a number of numerical and theoretical studies. Applied to {\it Herschel} data on the regions Aquila and Rosette, the method extracts pronounced power-law tails, consistent with those seen in simulations of evolved clouds. 
\end{abstract}


\section{Introduction}
The analysis of the probability density functions (pdf) of mass density (\rhopdf) and of column density (\Npdf) has been increasingly recognized as a key to understand the physics of star-forming regions. About two decades ago, it was shown from theoretical modelling \citep{VS_94, Passot_VS_98} that supersonic turbulence in an isothermal and non-gravitating medium produces a lognormal \rhopdf. This was confirmed by many numerical studies with varying characteristics of the environment like magnetic field, equation of state, Mach number, turbulence driving \citep[e.g.,][]{Padoan_Nordlund_Jones_97, Klessen_00, Li_Klessen_MacLow_03, Kritsuk_ea_07, Federrath_ea_10, Konstandin_ea_12, Molina_ea_12}. When the importance of gravity gradually increases in dense zones of the considered region, the \rhopdf~develops a power-law (PL) tail at its high-density end \citep{Klessen_00, Slyz_ea_05, VS_ea_08, Kritsuk_Norman_Wagner_11, Collins_ea_12, Federrath_Klessen_13}. In the course of further evolution of self-gravitating clouds, the point of deviation from the lognormal shape shifts to lower densities and the slope of the PL tail (hereafter, PLT) flattens towards a constant value \citep{Girichidis_ea_14}.

Could one discern similar signatures in the corresponding column-density distributions, derived from plane projections of the numerical cubes? Recent simulations of formation and contraction of molecular clouds (MCs) clearly hint at an evolution of the \Npdf, {\it morphologically analogous} to that of the \rhopdf. The obtained \Npdf s were found to have a lognormal shape in simulated turbulent clouds \citep{Federrath_Klessen_13, Ward_Wadsley_Sills_14, Matsumoto_Dobashi_Shimoikura_15} while in contracting/star-forming clouds their main part is still lognormal, but with a pronounced PLT  \citep{BP_ea_11, Kritsuk_Norman_Wagner_11, Federrath_Klessen_13, Auddy_Basu_Kudoh_18, Koertgen_ea_19}. In the latter case, there are asymptotic thresholds for slope and deviation point of the PLT at later evolutionary stages \citep{Ward_Wadsley_Sills_14}. Their values are not universal but depend on the sonic Mach numbers and the type of turbulence forcing \citep[see Fig. 4 in][]{Federrath_Klessen_13}. This general picture appears to be in good agreement with observations of star-forming clouds whose \Npdf s could be decomposed to a lognormal part and a PLT at their high-density ends \citep[e.g.,][]{Schneider_ea_13, Schneider_ea_15a, Pokhrel_ea_16}. Flattening of the slope with the cloud evolution seems also to be supported by the work of \citet{Abreu-Vicente_ea_15} who found shallow slopes for H {\sc ii} regions and much steeper -- for MCs at earlier stages of star formation. On the other hand, clear observational examples of a (quasi-)lognormal \Npdf{} for a quiescent cloud are only a few; for instance, in Polaris \citep{Schneider_ea_13} and in Draco and Spider \citep{Schneider_ea_19}.

In studies of a thermally bistable/multiphase interstellar medium, it has been found that the \rhopdf{} and the \Npdf{} have more complex shapes accounted for by various physical factors. Thermal instabilities in purely hydrodynamic flows may generate bimodal \rhopdf s while turbulent energy injections at small scales and/or including magnetic fields, Coriolis force and stellar feedback lead to evolutionary transition to a single-peaked distribution, possibly with a PLT \citep{VS_ea_00}. The type of turbulent forcing modifies additionally the high-density part of the \rhopdf{}: a PLT in the case of compressive forcing and a single lognormal shape, i.e. no bimodality, in the case of purely solenoidal forcing \citep{Seifried_Schmidt_Niemeyer_11}. \citet{BP_ea_11} demonstrated from numerical simulations of a thermally unstable gas that the \Npdf{} is also bimodal -- this is clearly visible at early evolutionary stages and still detectable at later stages, as a high-density PLT develops when self-gravity becomes important. These two cases are recently testified from {\it Herschel} observations of the diffuse cloud Draco \citep{Schneider_ea_19} and of MCs associated with \Hii regions \citep{Tremblin_ea_14}. In general, the emergence of a PLT in the (column-)density distribution is an established phenomenon in evolved thermally bistable/multiphase medium.

The decomposition of \Npdf s in star-forming clouds in to a (quasi-)lognormal part and a PLT is also a challenge from the perspective of observations. It has been currently debated whether a lognormal part can be extracted reliably. Indeed, the very low column-density range is affected by noise biases, uncertainties due to map area and small dynamic range of dust observations while the lognormal fit remains the best model for this part of the \Npdf{} \citep{Ossenkopf_ea_16, Chen_ea_18}. \citet{Lombardi_Alves_Lada_15} and \citet{Alves_Lombardi_Lada_17} put to question the detectability of a lognormal part of observational \Npdf s at all,  attributing seemingly lognormal shapes at low column densities to incompleteness of data and sampling effects (`last closed contour'). However, \citet{Koertgen_ea_19} demonstrate from magnetohydrodynamical simulations, that in a turbulently mixed medium there is no `last closed contour' for a sufficiently low column-density threshold.

On the other hand, \citet{Brunt_15} question whether a PL fit is indeed an appropriate description of the high column-density part. He argues that \Npdf s, derived on the base of dust-extinction maps of MCs, could be reproduced combining two lognormals that account respectively for the warm and the cold gas included the selected region.     

But even taking for granted that the \Npdf~in a star-forming cloud consists of two clearly recognizable parts, there is another, {\it methodological} issue. The characteristics of the two fitting functions are obviously interdependent. For instance, the usual procedure to extract the PLT is as follows: 
\begin{enumerate}
 \item Find the best lognormal fit of the main (low-density) \Npdf~part (e.g. by $\chi^2$ goodness).
 \item Estimate the `deviation point' (DP) of the distribution from the lognormal fit (e.g. by  $3\sigma$ criterion, with $\sigma$ being the Poissonian data uncertainty in the considered bin).
 \item Fit the rest of the distribution with a PL function.
\end{enumerate}

Such approach rests on the assumption that the main \Npdf~part is lognormal and thus the resulting DP and the PL slope depend on the lognormal parameters. This is problematic not only in view of the abovementioned debates on lognormality of the \Npdf~at lower densities. If the PL regime is to be interpreted as a signature of the impact of self-gravity, then the slope value is an indicator of their evolutionary stage. It has been demonstrated that the PL slope is interrelated to the density profile of spherically symmetric, collapsing, isothermal clouds \citep{Kritsuk_Norman_Wagner_11} and can be representative for the general structure of real MCs \citep{Donkov_Veltchev_Klessen_17}. Therefore one needs a method to extract the PLT on minimal assumptions about the rest of the column-density distribution.

In this paper we propose an objective method to extract the PLT of the \rhopdf~and \Npdf{} in star-forming MCs. The method is based on the statistical approach to test the power-law features of an arbitrary distribution, developed by \citet{Virkar_Clauset_14}, as the PL slope and the DP are estimated simultaneously. In Section \ref{Method} we discuss the original approach and comment on its applicability on large datasets. Section \ref{Data} reviews the numerical and observational data used in this work. The proposed method itself is described in Section \ref{Results: Adapted method}. The results from its application to (column-)density distributions from simulations and in two Galactic star-forming regions are presented in Sections \ref{Results: Application to numerical data} and \ref{Results: application to Herschel data}, respectively. We conclude this work with discussion (Sect. \ref{Discussion}) and summary (Sect. \ref{Summary}).

\section{Mathematical method}       \label{Method}
\subsection{Application to unbinned data} \label{The PLFIT method}

The original statistical approach ``to discern and quantify power-law behaviour in empirical data'' was proposed by \citet{Clauset_Shalizi_Newman_09}. It deals with {\it unbinned} observational data and thus avoids possible subjectivity of the PL fit introduced through the choice of binning size. Assuming that the data above a chosen lower limit $x_{\rm min}$ follow a power law $p(x) \propto x^{-\alpha}$, the method derives an estimate of this limit $\hat{x}_{\rm min}$ through minimization of the Kolmogorov-Smirnov (KS) goodness-of-fit statistics and of the slope $\hat{\alpha}=\hat{\alpha}(\hat{x}_{\rm min})$ through maximum-likelihood estimators (see \citealt{Clauset_Shalizi_Newman_09} for details). The procedure does not rule out that other (non-PL) functions might better fit the observed distribution -- it derives the range and the slope of the best possible PL fit. 

A software implementation of the method, called {\sc Plfit}\footnote{http://tuvalu.santafe.edu/$\sim$aaronc/powerlaws/}, has been applied for PLT extraction from \Npdf s in zones of the Perseus star-forming region \citep{Stanchev_ea_15}. Unfortunately, the use of this program is technically limited to smaller datasets. 

\subsection{Application to binned data}  \label{The bPLFIT method}
To deal with large datasets of size $\gtrsim 10^5$ points from numerical simulations and high-resolution imaging of MCs, one could use a version of the approach, described in the previous subsection, adapted for {\it binned} distributions (\citealt{Virkar_Clauset_14}; software implementation is also available\footnote{http://tuvalu.santafe.edu/$\sim$aaronc/powerlaws/bins}). The technique does not depend on assumptions about the binning scheme and is applicable to linear, logarithmic and arbitrary bins. Hereafter, we call this method \bPLFIT. 

Like its unbinned counterpart, \bPLFIT{} uses maximum-likelihood estimators to assess the PLT. The required input is a binned distribution $(\{b_i\}, 1\le i\le k+1; \{h_i\}, 1\le i\le k)$ where $b_i$ are the bin boundaries and $h_i$ are the corresponding counts for $k$ bins. The approach is realized in several basic steps:
\begin{enumerate}
 \item[1.] Choose consecutively test values $b_{{\rm min,}\,i}$ for the DP, from the set $(b_1, ... ,b_k)$.
 \item[2.] Calculate the slope $\alpha_i (b_{{\rm min,}\,i})$ for each test DP value. 
 \item[3.] Compute the KS goodness-of-fit statistic for each $(\alpha_i, b_{{\rm min,}\,i})$.
 \item[4.] Find the PL parameters $(\hat{\alpha}, \hat{b}_{\rm min})$ that minimize the KS statistic.
\end{enumerate}

In case $b_{\rm min}$ is set by the user to some constant value, the method yields a single corresponding slope $\alpha$.

Using synthetic data, the authors of the method performed a number of tests of the PL hypothesis against alternative distributions like lognormal, exponential and a few others. Using the log-likelihood ratio as criterion, they found that the PL distribution can be correctly distinguished from lognormal for dataset sizes as large as $\sum_{i=1}^k h_i \gtrsim 2\times 10^4$ and bin-sizes $b_{i+1}/b_i \le 2$ (\citealt{Virkar_Clauset_14}; Sect. 5). 

Therefore \bPLFIT{} can be adopted as a reliable approach to extract PLTs from \rhopdf s or \Npdf s given that the datasets are large enough. The critical issue is how to minimize the possible subjectivity introduced through the choice of total number of bins $k$. We develop a procedure to derive average PLT parameters, using logarithmic binning. The procedure will be presented in Section \ref{Results: Adapted method} and then applied to simulated data on evolving, self-gravitating MCs (Section \ref{Results: Application to numerical data}) and on {\it Herschel} data for two Galactic regions of active star formation (Section \ref{Results: application to Herschel data}). 

\section{Used numerical and observational data}
\label{Data}

We make use of data from two sets of simulations (Sec. \ref{SILCC data} and \ref{HRIGT data}) and of dust-continuum observations by {\sl Herschel} (Sec. \ref{Herschel data}). The simulations were performed using the hydrodynamical code \textsc{Flash} \citep{FLASH00,Dubey2008}. The first numerical data set consists of a magneto-hydrodynamical (MHD) simulations from the SILCC project (Section \ref{SILCC data}), including feedback effects and tracing the evolution of giant MCs in a galactic environment. The scales cover range from 500\,pc down to 0.12\,pc. The second set are isothermal hydrodynamical (HD) simulations of self-gravitating clouds with supersonic turbulence at scales of typical large clumps ($0.5$~pc) in MCs, with adaptive resolution going as high as $\sim3$~au in high-density zones. The very high resolution allows to resolve the high-density end of the PLT; see Section \ref{HRIGT data} for details. Hereafter, this set of simulations is called HRIGT (High-Resolution Isothermal Gravo-Turbulent).

The observational data stem from {\sl Herschel} imaging that provides dust column-density maps at angular resolutions lower than 1$'$. \Npdf s obtained from these maps were presented for a large number of molecular cloud types, ranging from quiescent regions \citep{Schneider_ea_13} to low- and high-mass star-forming regions \citep[e.g.][]{Russeil_ea_13, Konyves_ea_15, Schneider_ea_15a, Schneider_ea_15b, Stutz_Kainulainen_15, Rayner_ea_17}.

\subsection{Galactic-scale (SILCC) simulations} \label{SILCC data}
The simulation framework including the physical processes is based on the SILCC project \citep{SILCC1, SILCC2}. The details of the particular setup as well as the dynamical evolution are described in \citet{SeifriedEtAl2017} and \citet{SILCC5}, so we only highlight the most important aspects of the simulations here. We set up a stratified gas distribution in a simulation box of $0.5\times0.5\times\pm5\,\mathrm{kpc}^3$. The total gas surface density is $\Sigma_\mathrm{gas}=10\,\mathrm{M}_\odot\,\mathrm{pc}^{-2}$, which results in a total mass of $2.5\times10^6\,\mathrm{M}_\odot$ in the box. The gas initially follows a Gaussian distribution with a scale height of $30\,\mathrm{pc}$. 
Gravitational effects are included in two ways. We account for the stellar component of the galaxy using an external potential, which is implemented as an isothermal sheet with a stellar surface density of $30\,\mathrm{M}_\odot\,\mathrm{pc}^{-2}$. The second component of the gravitational force is self-gravity, which is computed with a tree-based solver \citep{WuenschEtAl2018}. Radiative cooling is computed via a non-equilibrium chemical network \citep[see e.g.][for reference]{Glover_ea_10} that actively follows the abundances of ionized (H$^+$), atomic (H) and molecular (H$_2$) hydrogen as well as carbon monoxide (CO) and singly ionized carbon (C$^+$). The gas is initially purely atomic and at rest.

We let supernovae (SNe) explode at a constant rate of $15\,\mathrm{Myr}^{-1}$. The latter is derived using the Kennicutt-Schmidt relation \citep{Schmidt1959, KennicuttSchmidt1998} and assuming the initial mass function by \citet{Chabrier2003}. We follow the evolution of the gas for about $50\,\mathrm{Myr}$ at a maximum resolution of $\Delta x=3.9\,\mathrm{pc}$ in order to identify regions in which molecular clouds form. We then restart the simulations at $t=11.9\,\mathrm{Myr}$ and progressively increase the adaptive refinement up to a maximum resolution of $\Delta x=0.12\,\mathrm{pc}$ while ensuring that the Jeans length is always refined with at least 16 cells. The resolution needs to increase gradually as described in detail in \citet{SeifriedEtAl2017}. The simulations do not include sink particles but focus on the evolution of the gas phase.

\subsection{Clump-scale (HRIGT) simulations}  \label{HRIGT data}

The second set of simulations focuses on smaller scales of star-forming regions. For this purpose we set up cubic periodic boxes with a length of $0.5\,\mathrm{pc}$. We reduce the complexity in physical modeling in favour of higher resolution and significantly higher adaptive refinement. We therefore omit the details of chemical phases and local shielding effects. The gas is isothermal at a temperature of $10\,\mathrm{K}$. We assume a mean molecular weight of 2.3, which results in an isothermal sound speed of $0.19\,\mathrm{km\,s}^{-1}$. We set up a homogeneous gas distribution with two mean densities $\langle\rho\rangle$ ($4.6\times10^{-20}\,\mathrm{g\,cm}^{-3}$ and $2.3\times10^{-19}\,\mathrm{g\,cm}^{-3}$) corresponding to a total mass of $85$ and $426\,\mathrm{M}_\odot$, respectively. Due to the initially uniform density and the isothermal temperature we can express the total mass in units of the initial average Jeans mass $M_\mathrm{J}=\langle\rho\rangle(\pi/6) \lambda_\mathrm{J}^3$, where $\lambda_\mathrm{J}=(\pi c_\mathrm{s}/G\langle\rho\rangle)$ with the isothermal sound speed $c_\mathrm{s}$. For our setup the Jeans masses are $M_{\mathrm{J},\,0}=2.7\,\mathrm{M}_\odot$ and $M_{\mathrm{J},\,0}=1.2\,\mathrm{M}_\odot$ for the two different densities. The total mass in the box is then $32$ and $354$ Jeans masses. The initial turbulent velocities are constructed in Fourier space with a peak of the power spectrum at $k=2$, i.e. half of the box size. The spectral index for the velocity is set to $-2$ representing Burgers turbulence for strong shocks. We distinguish between purely compressive, purely solenoidal and naturally mixed velocities \citep{Federrath_Klessen_Schmidt_08}. The first case corresponds to radial projections in Fourier space, the second to tangential projections, and for the mixed case we do not apply any projection, which statistically results in approximately one third of the power in compressive and the remaining two thirds in solenoidal motions \citep{Federrath_Klessen_Schmidt_08}. We would like to emphasize that the nature of turbulent modes (compressive vs. solenoidal) has a strong impact on the time scales needed to trigger the formation of clouds and the onset of gravitational collapse in over-densities, see table~\ref{table_HRIGT_runs}. We only set the initial motions and let the turbulence decay. The root-mean-square velocity of the gas is set to $0.38\,\mathrm{km\,s}^{-1}$, i.e. the initial gas motions are supersonic with a Mach number of $\mathcal{M}=2$. This corresponds to a crossing time of approximately $1.3\,\mathrm{Myr}$. We start with a uniform grid of $256^3$ cells and progressively refine up to an effective resolution of $32768^3$ cells, such that the Jeans length is always refined with at least 16 cells. At the highest densities the cells thus have a size of $3.15\,\mathrm{au}$. Self-gravity is included from the beginning of the simulations, which causes the the gas to become locally unstable and collapse. At a threshold density of $\rho=2.12\times10^{-13}\,\mathrm{g\,cm}^{-3}$ we introduce Lagrangian sink particles based on \citet{Federrath2010}.
 
\begin{table}
\caption{HRIGT runs selected to elaborate the method for PLT extraction. Notation: v1, v2, v3 -- chosen velocity field; vfmode -- velocity field mode; m(ixed), c(ompressive), s(olenoidal) -- acronyms for the vfmode (see text); $M_{{\rm J},\,0}$ -- initial Jeans mass in the box; $\tau_{\rm ff}(\langle\rho\rangle)$ -- free-fall time of the mean density; $t_{\rm sim}$ -- run duration.}
\label{table_HRIGT_runs} 
\begin{center}
\begin{tabular}{lc@{~~}c@{~}c@{~~~}c@{~~}c}
\hline 
\hline 
Name & \multicolumn{2}{c}{Total mass}  & vfmode & $\tau_{\rm ff}(\langle\rho\rangle)$ & $t_\mathrm{sim}$ \\ 
~    & [$M_\odot$] & [$M_{{\rm J},\,0}$] & ~ & [ kyr ] & [ $\tau_{\rm ff}(\langle\rho\rangle)$ ] \\
\hline 
v1m-M085 & ~85 & ~32 &	mixed       & $310$ & $1.05$ \\
v1c-M426 & 426 & 354 &	compressive & $139$ & $0.41$ \\	
v1s-M426 & 426 & 354 &	solenoidal  & $139$ & $1.57$ \\
v2m-M085 & ~85 & ~32 &	mixed       & $310$ & $2.50$ \\	
v2m-M426 & 426 & 354 &	mixed       & $139$ & $0.80$ \\
v3m-M085 & ~85 & ~32 &	mixed       & $310$ & $1.38$ \\
\hline 
\hline 
\end{tabular} 
\end{center}
\smallskip 
\end{table}

\begin{figure*} 
\begin{center}
\includegraphics[width=.9\textwidth, keepaspectratio]{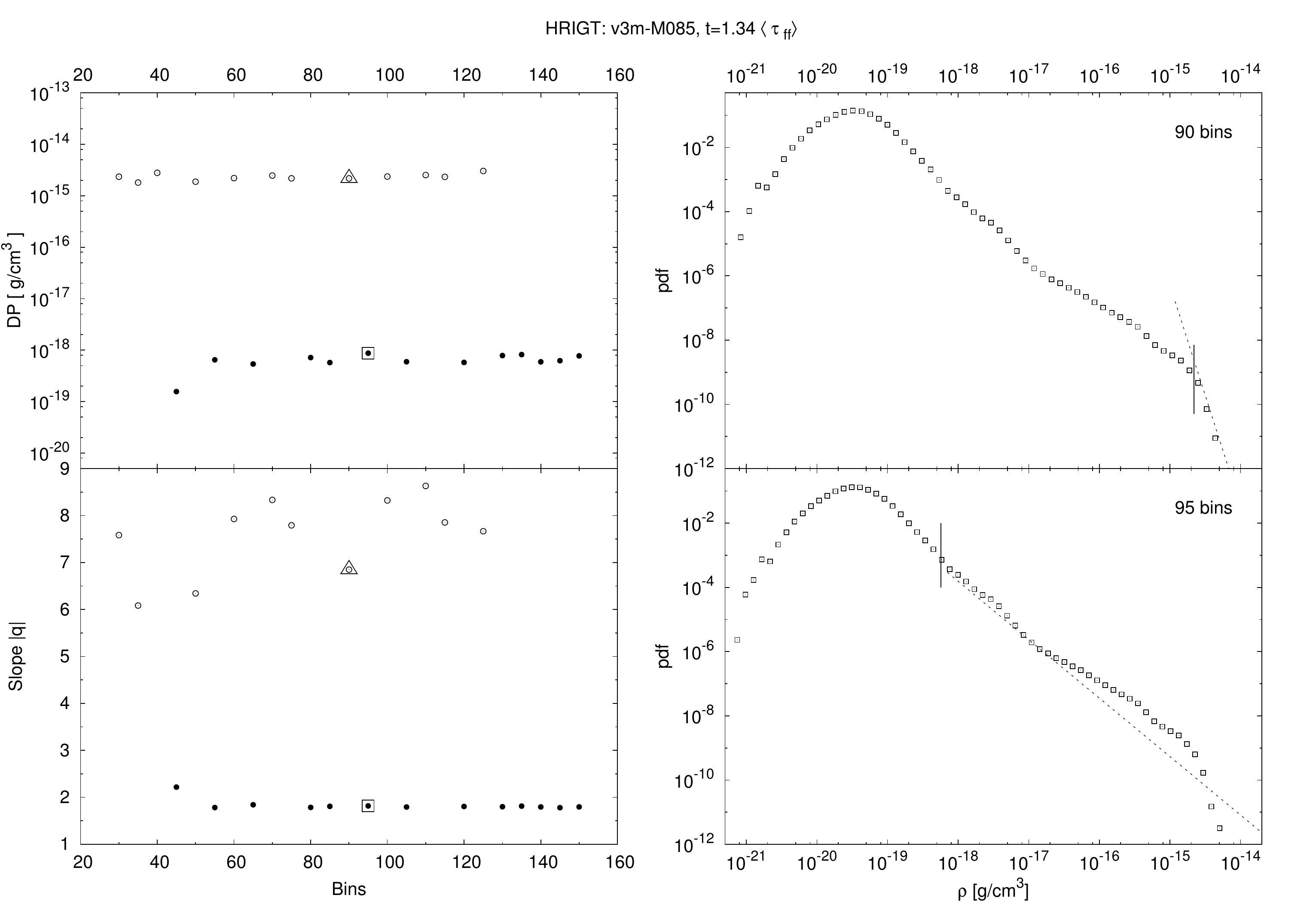}
\caption{On the dependence of \bPLFIT{} output from the chosen number of logarithmic bins $k$. The HRIGT run v3m-M085 is selected for illustration. DPs (top) and slopes (bottom) of the \rhopdf{} PLT as functions of $k$ are plotted in the left column. The extracted short ($\le 5$ bins) and long tails are indicated with open and filled circles, respectively. The open triangle denotes an example of \rhopdf{} with extracted short tail (top right) while open square stands for a \rhopdf{} with long tail (bottom right). In the right panels, short solid lines indicate the DP estimates and the derived slopes are plotted with dotted lines.}
\label{fig_binning_dependence}
\end{center}
\end{figure*}

\subsection{{\it Herschel} observations of star-forming regions} \label{Herschel data}
We produced dust column-density maps of the Aquila and Rosette regions from publicly available level-3 data products produced with HIPE 13 from the {\it Herschel} archive, using the observations at 160, 250, 350 and 500 $\mu$m. Aquila was observed within the Gould Belt keyprogram \citep{Andre_ea_10} and presented in \citet{Bontemps_ea_10}, \citet{Konyves_ea_10, Konyves_ea_15} and \citet{Schneider_ea_13}. Rosette was part of the HOBYS keyprogram \citep{Motte_ea_10} and is discussed in \citet{Schneider_ea_10}, \citet{Schneider_ea_12}, \citet{Hennemann_ea_10} and \citet{DiFrancesco_ea_10}.

The angular resolution of the maps is 11\das7, 18\das2, 24\das9, and 36\das3 for 160, 250, 350, and 500 $\mu$m, respectively.  For an absolute calibration of maps (included in the SPIRE level-3 data), the {\it Planck} -HFI observations were used for the HIPE-internal {\tt zeroPointCorrection} task that calculates the absolute offsets, based on cross-calibration with HFI-545 and HFI-857 maps, including colour-correcting HFI to SPIRE wavebands, assuming a grey-body function with fixed $\beta$. The zero-point determination for the PACS 160 $\mu$m map was performed using IRAS data, following the procedure described in \citet{Bernard_ea_10}. Column density and temperature maps were then produced at an angular resolution of 18\arcsec, following the procedure outlined in \citet{Palmeirim_ea_13} which employs a multi-scale decomposition of the imaging data and assuming a constant line-of-sight temperature. We performed a pixel-by-pixel SED fit from 160 to 250 $\mu$m, using a dust opacity law $\kappa_0 = 0.1 \times (\nu/1000~{\rm GHz})^\beta$ cm$^2$ g$^{-1}$ which is similar to that of \citet{Hildebrand_83} adopting $\beta = 2$ and assuming a gas-to-dust ratio of 100. (This dust opacity law is commonly adopted in other papers on the Gould Belt survey and HOBYS.) We estimate that the final uncertainties of the column-density map are around 20 -- 30 \%. 

We selected the Aquila and Rosette molecular clouds because they represent examples of molecular clouds forming mostly stars of low and high mass, respectively.  The area of the Aquila cloud that was covered with {\sl Herschel} imaging is the western high extinction region of the Aquila Rift that extends over $\sim$20$^\circ$ located above the Galactic plane (b$\sim$4$^\circ$). It contains the young star cluster W40, associated with the H\,{\sc ii} region Sharpless 2-64.  The molecular mass of the Aquila cloud determined from the {\sl Herschel} column density map within a radius of 9.8 pc is $\sim2\times10^4$ M$_\odot$. The mass value assumes a distance of 260 pc \citep{Straizys_ea_03} for the entire Aquila complex, though a larger distance of around 415 pc \citep{Dzib_ea_10} is also discussed in the literature and recently confirmed by {\it Gaia} measurements to be 439 pc \citep{Ortiz-Leon_ea_18}. The Rosette molecular cloud is an evolved region that already formed a cluster of 17 OB stars (NGC 2244), located in the center of the cloud. Nevertheless, there is continuous star formation along the central molecular ridge of the Rosette cloud \citep{Hennemann_ea_10, Schneider_ea_12}.  The total mass of the complex from the {\sl Herschel} column density map is $\sim$10$^5$ M$_\odot$, assuming a distance of 1.6 kpc \citep[e.g.][]{Turner_76}. With the adopted distance estimates, the linear resolutions of the maps are 0.04 pc for Aquila (1 pc$=$7\das 9) and 0.14 pc for Rosette (1 pc$=$2\das 15). The column-density maps are provided in Appendix \ref{Maps of the SF regions} for reference.

\section{Adapted method to extract PLTs from pdfs}   \label{Results: Adapted method}
\subsection{Criterion for plausible PLT extraction}
\label{Criterion_for_plausible_PLT_extraction}
As \bPLFIT{} is applied to a given density distribution, one asks how does the result depend on the chosen total number of bins $k$. Fig. \ref{fig_binning_dependence} (left) shows the \bPLFIT{} outputs with $k$ varied in a large range, taking an example from the HRIGT runs. One clearly distinguishes two subsets of extremely different slopes/DPs. Within them, both PLT parameters do not seem to correlate with $k$. Examples of pdfs from each subset are shown in Fig. \ref{fig_binning_dependence}, right (denoted by open square and open triangle in the left panel). Although differences between the plotted distributions are hardly discernable by eye, \bPLFIT{} extracts a long, shallow PLT from the pdf displayed in the bottom-right panel and a very short, steep PLT from the pdf in the top-right panel. In the former case, the derived slope is not affected by the drop of statistics close to the upper data limit. In the latter case, the PL fit is based exactly on the last few bins; thus it reflects the incompleteness of data and should be considered as false. Inspection of the \bPLFIT{} outputs from all sampled numerical and observational data (Sections \ref{SILCC data}--\ref{Herschel data}) showed that such false PLT extractions occur in an unpredictable way, for various choices of $k$; sometimes they alternate with extractions of long PLTs as $k$ is being increased (see Fig. \ref{fig_binning_dependence}, left). This is a peculiarity of the \bPLFIT{} technique itself and can not be overcome, e.g., by introducing some cut-off of the pdf at the high-density end. The simplest and most straightforward solution of the problem is to exclude false PLT extractions from further consideration. Hence a necessary first step for plausible PLT extraction should be to select pdfs with large enough PLT spans in units of bins $b_{k+1}/b_{\rm min}>{\tt FALSE}$ where ${\tt FALSE}$ is the maximal span of false PLTs. The latter typically is only two\footnote{Since two points determine a straight line (a PL function on a log-log plot) and \bPLFIT{} "fits" them in case a larger PLT is not recognized.} or three bins; thus ${\tt FALSE}=5$ bins would be a conservative and secure choice. PLTs that pass the false-extraction test (filled symbols in Fig. \ref{fig_binning_dependence}, left) can be sampled to calculate averaged DP and PL slope in the considered density distribution.

\begin{figure} 
\begin{center}
\begin{tabular}{c}
\includegraphics[width=84mm]{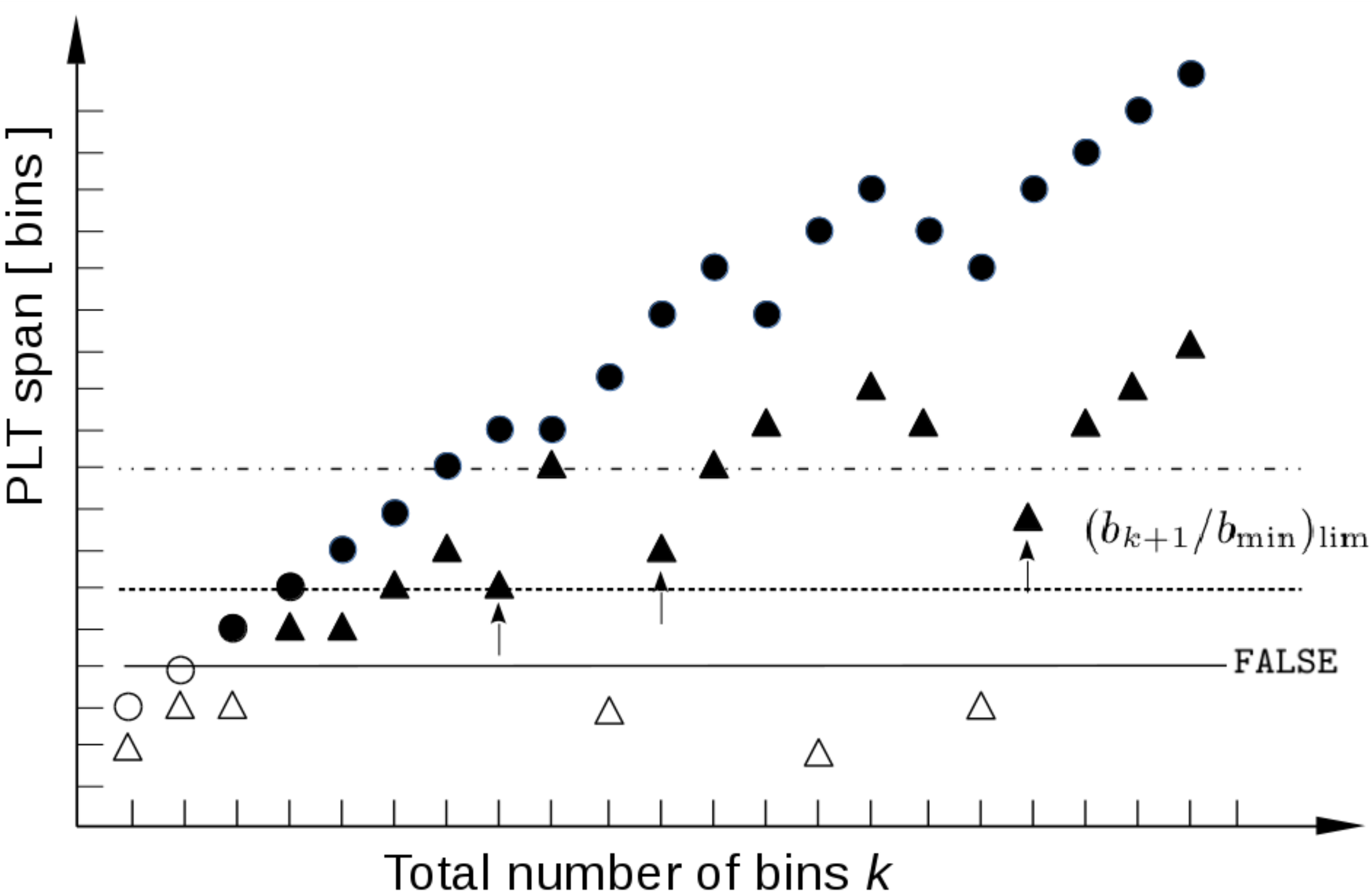}\\
\includegraphics[width=84mm]{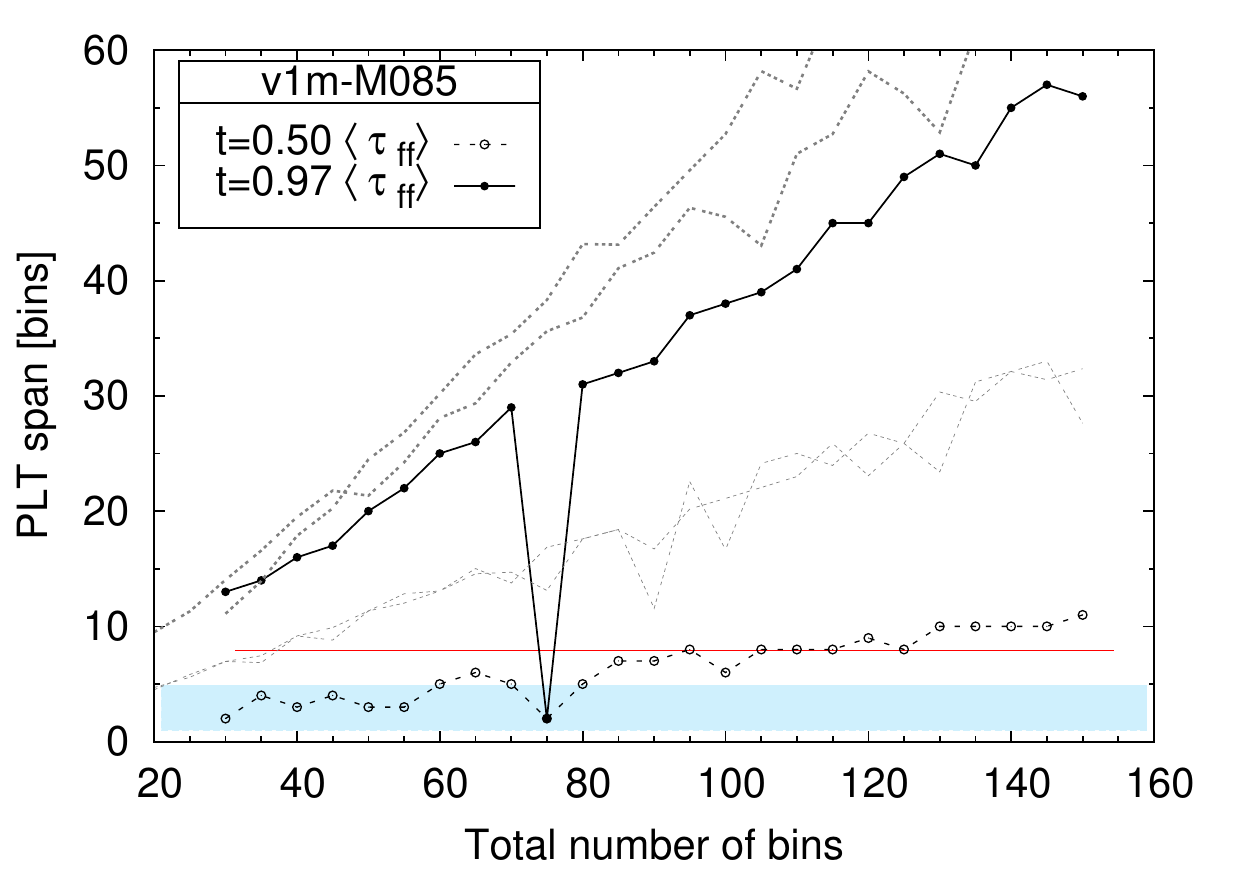}
\end{tabular}
\caption{Span of the detected PLT as a function of the total number of bins. {\it Top:} Illustration on the cases of rich (circles) and poor (triangles) statistics in the supposed PLT range. Open symbols denote false PLTs with spans $\le {\tt FALSE}$ while artificially shorter but plausible PLT extraction are marked with arrows. Two choices of a $(b_{k+1}/b_{\rm min})_{\rm lim}$ are shown (dash-dotted and dotted lines). The choice of the lower value provides larger samples of plausible PLT detections. {\it Bottom:}  On the criterion for sampling of pdfs with reliable PLT parameters. Two examples of \bPLFIT{} outputs from the HRIGT run are selected for illustration. They are compared with analytically defined pdfs (grey dashed lines) with approximately the same PLT slopes: $\alpha=-1.9$ (thick) or $\alpha=-3.6$ (thin). (See main text, for further details.) The shaded area denotes the zone of binned distributions that fail the false-extraction test with ${\tt FALSE}\equiv 5$~bins; the adopted choice of $(b_{k+1}/b_{\rm min})_{\rm lim}=8$~bins is plotted with a red line. }
\label{fig_PLT_range-bins}
\end{center}
\end{figure}

Second, one needs to quantify the minimal PLT span $(b_{k+1}/b_{\rm min})_{\rm lim}$ (hereafter, {\it lower plausibility limit}) which defines a plausible \bPLFIT{} output in terms of fit's goodness. The choice of $(b_{k+1}/b_{\rm min})_{\rm lim}$ should provide a sufficient number of plausible PLT extractions to calculate an average DP and slope. The total statistics within the supposed PLT density range is important here, as illustrated in Fig. \ref{fig_PLT_range-bins}, top. In the case of rich statistics (circles), i.e., long and shallow tails, the PLT span increases quasi-monotonically with the chosen total number of bins $k$. Adopting a high value of $(b_{k+1}/b_{\rm min})_{\rm lim}$ (dash-dotted line) is affordable -- it corresponds to some minimal $k$ above which all PLT extractions are plausible. However, in case of poor statistics (triangles), i.e., short, unevolved tails, the relation between $k$ and the PLT span is more complicated. The latter could be underestimated due to emergence of `wiggles' or `spikes' as $k$ is increasing (bin size is decreasing). Artificially shorter (but still plausible) tails can be detected for some choices of $k$ (shown with arrows in Fig. \ref{fig_PLT_range-bins}, top) due to a relatively low pdf value in the bin, adjacent from left to the estimated DP. Then one needs to choose a lower $(b_{k+1}/b_{\rm min})_{\rm lim}$ (dotted line) to get a larger sample of detected plausible PLTs for derivation of averaged \bPLFIT{} outputs. 

Fig. \ref{fig_PLT_range-bins}, bottom, shows the results from tests of the criterion for plausible PLT extraction. We selected two evolutionary stages of a HRIGT run. At $t=0.5 \langle \tau_{\rm ff} \rangle$ the PLT is still emerging, with steep extracted slopes ($-3 > \alpha \gtrsim -6$). Its span is small and one needs to adopt a lower value of $(b_{k+1}/b_{\rm min})_{\rm lim}$ (red line) to obtain a richer sample of PLT extractions. At an evolutionary stage as late as $0.97 \langle \tau_{\rm ff} \rangle$  the PLT span $b_{k+1}/b_{\rm min}$ increases fast and almost monotonically with the bin size, except one false PLT extraction for $k=75$~bins. In this case, one could adopt a lower plausibility limit up to three decades. 

For comparison, we applied \bPLFIT{} on analytical pdfs that are combinations of a lognormal function and a PLT and are good approximations of the pdfs from the chosen HRIGT run. In contrast to distributions from numerical/observational data, analytical pdfs are smooth and possess a bin-independent statistics\footnote{The pdf value in given bin always reflects the real distribution, regardless of how small the bin size is.}. The results for two pairs of analytical pdfs are plotted in Fig. \ref{fig_PLT_range-bins}, bottom (grey dashed lines). Thin lines indicate steep PLTs ($\alpha=-3.6$), corresponding to $t=0.50 \langle \tau_{\rm ff} \rangle$, and thick lines denote shallow PLT slopes ($\alpha=-1.9$), corresponding to $t=0.97 \langle \tau_{\rm ff} \rangle$. As expected, the PLT span of the analytical pdfs behaves similarly to that of the numerical distributions but without detections of false tails. The \bPLFIT{} method extracts correctly the slope, with vanishing deviations from the real value, while the typical deviations of the DP are of order of the bin size (see Appendix \ref{Appendix_bPLFIT_analytical pdfs}). 

\subsection{Average PLT parameters}
Now we are able adopt a procedure for  extraction of PLT with an average span and slope, by use of \bPLFIT{}. The steps are as follows:
\begin{enumerate}
\item Choose the range of total bin number $k$. (Normally, $k$ could be varied from a few dozens to about two hundred, depending on the given dataset.)
\item Construct the pdf for each choice of $k$. 
\item Run \bPLFIT{} on each pdf and get the PLT parameters (slope and DP).
\item Set a maximal span of false PLTs ${\tt FALSE}$ in units of bin size and exclude false PLT detections from further consideration. (Recommended conservative choice: ${\tt FALSE}=5$ bins.) 
\item Set the lower plausibility limit  $(b_{k+1}/b_{\rm min})_{\rm lim}$ in units of bin size for the PLT span, depending on the statistics in the supposed PLT range. 
\item Select a sample of plausible PLTs with spans $(b_{k+1}/b_{\rm min}) \ge (b_{k+1}/b_{\rm min})_{\rm lim}$.
\item Calculate average slope and DP from the sample. 
\end{enumerate}

This adapted \bPLFIT{} method is appropriate for extraction of PLTs from $\rho$- or \Npdf s in star-forming clouds.
In the next Section we apply it to the selected data from numerical simulations as well to column-density data on two Galactic regions. 

\section{Application of the method to numerical data}    
\label{Results: Application to numerical data}
In this Section we present the results from application of the adapted \bPLFIT{} method to the selected simulation runs. We define logarithmic density $s=\log(\rho/\rho_0)$ and logarithmic column-density $z=\log(N/N_0)$, using some arbitrary normalization units. The PLTs of the \rhopdf{} and the \Npdf{} are introduced as follows:
\begin{eqnarray}
p(s)&=&A_{s}(\rho/\rho_0)^q \\
p(z)&=&A_{z}(N/N_0)^n~,
\end{eqnarray}
where $A_{s}$ and $A_{z}$ are constants. The adopted lower plausibility limit $(b_{k+1}/b_{\rm min})_{\rm lim}$ for the PLT span is 8 bins, both for the SILCC and the HRIGT data. 

\begin{figure*} 
\begin{minipage}{\textwidth}
\includegraphics[width=84mm]{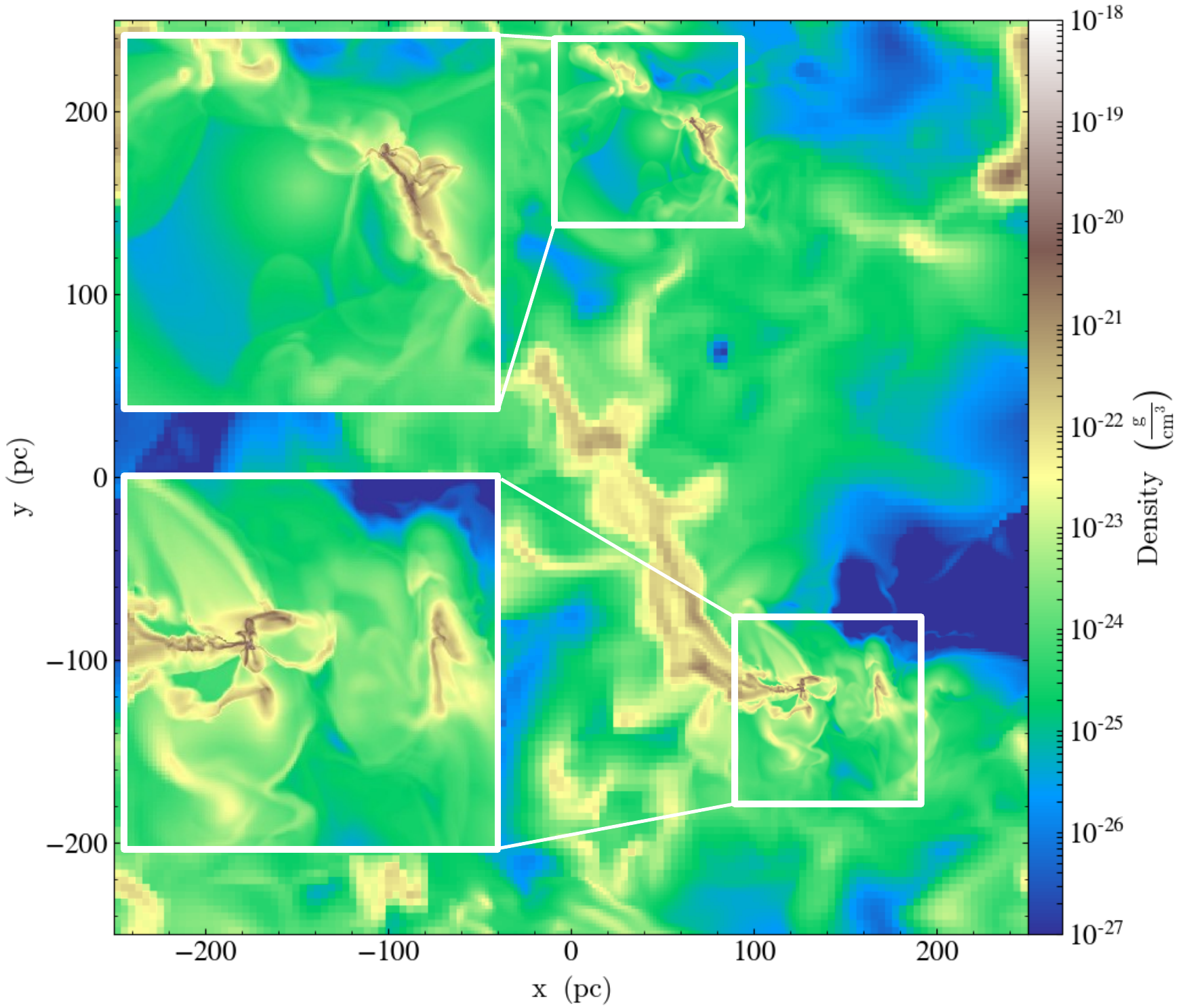}
\includegraphics[width=84mm]{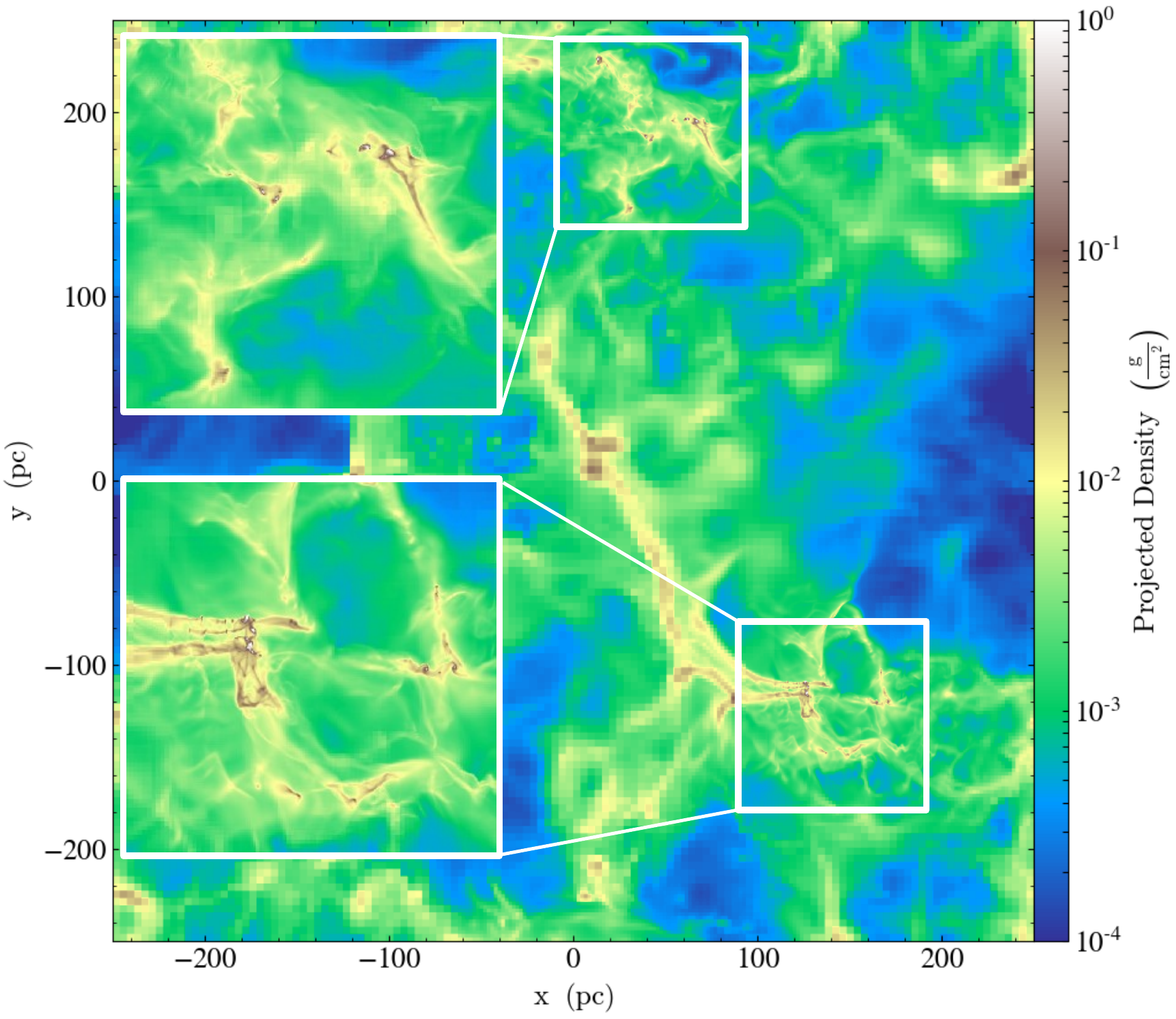}
\caption{Snapshot of density (left) and column-density (right) field at the end of the SILCC run. Two zones of higher resolution, containing evolved GMCs are zoomed in. The slice of the density field is taken at the galactic plane ($z=0$). We highlight the high-resolution regions, which mainly contribute to the high-density part of the PDF.}
\label{fig_SILCC_density_maps}
\end{minipage}
\end{figure*}

\begin{figure*}
\begin{center}
\includegraphics[width=1.\textwidth]{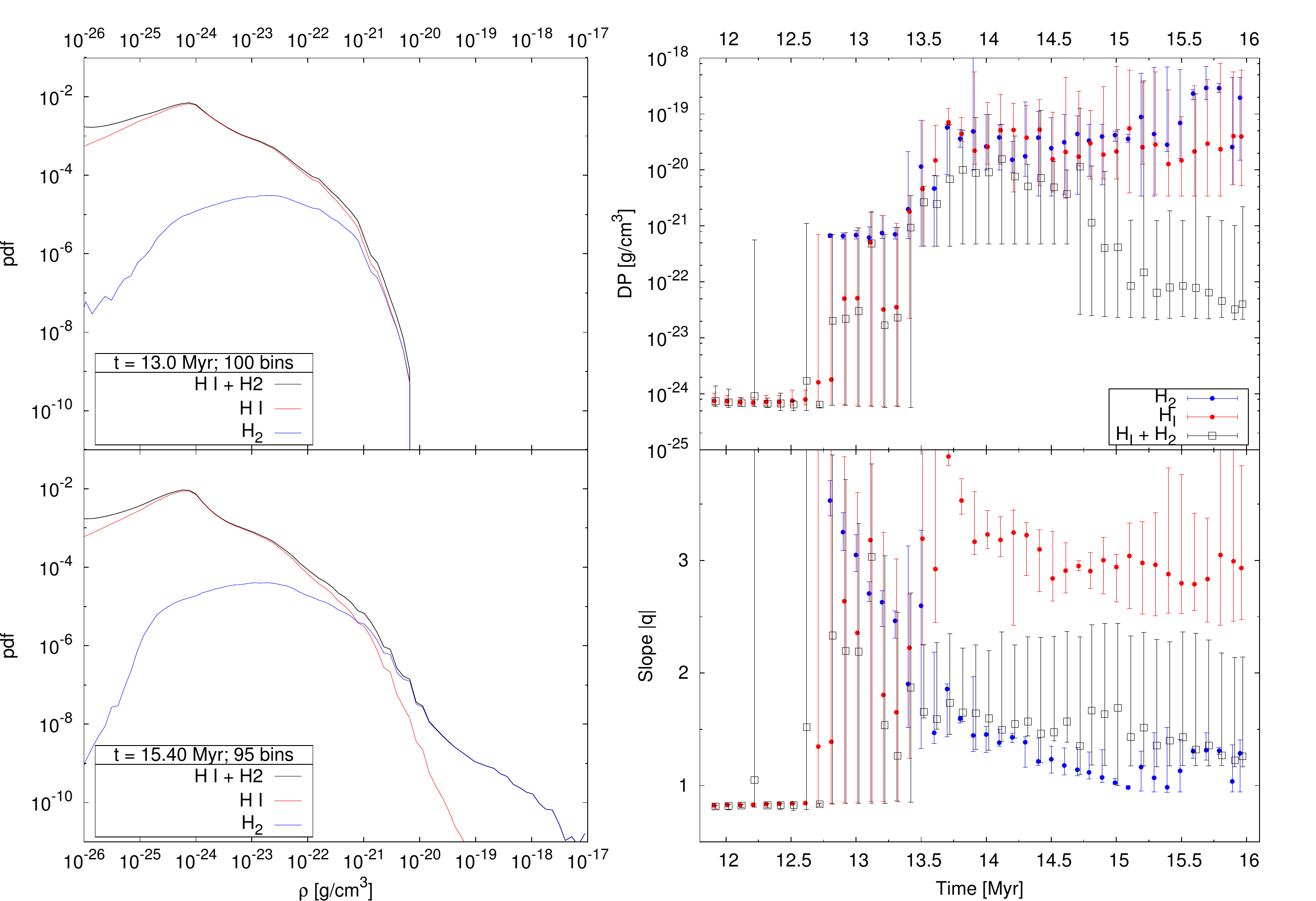}
\caption{Evolution of the \rhopdf~PLT from the SILCC data. {\it Left:} \rhopdf s at the evolutionary stages with an emerging (top) and well developed (bottom) PLT; {\it Right:} Evolution of the average PLT parameters. The error bars indicate the minimal/maximal value in the sample.}
\label{fig_rho-pdf_evolution_SILCC}
\end{center}
\end{figure*}

\subsection{Pdf evolution on galactic scales}
\label{Galactic-scale_pdf_evolution}
The chosen time span of the SILCC run is between $12$ and $16$~Myr, i.e. from the stage of emergence of high-density, self-gravitating clouds to the stage of formation of the first stars in them. The density and column-density distributions at the end of the run are shown in Fig. \ref{fig_SILCC_density_maps}. Here, we would like to highlight the two zoom region with higher resolution because they contribute mostly to the high density PLT. Therein one clearly identifies dense ($\rho\gtrsim 10^{-22}$~\gcc), elongated and filamentary structures with size dozens of pc. They can be considered as giant star-forming clouds at an early stage of their evolution. High-density cores within them are visible in the zoomed-in frames in Fig. \ref{fig_SILCC_density_maps}; they display sizes of several parsec and $N\gtrsim 10^{-2}$\gsqc~($3\times10^{21}$~\sqc), typical for molecular clouds. 

By applying the adapted \bPLFIT~to the entire SILCC cube, we study the evolution of the \rhopdf~of the gas and of its atomic and molecular phases separately. Fig. \ref{fig_rho-pdf_evolution_SILCC} (right) presents the results in terms of the PLT parameters as the total number of bins is varied in the range $15\le k\le 150$. In the very early phase ($t\lesssim 12.8$~Myr) the atomic gas dominates and its \rhopdf~displays a shallow PLT with DP in the diffuse regime ($\sim 10^{-24}$~\cc), i.e. not attributable to gravitational contraction (see Fig. \ref{fig_PLT_early_evolution} in Appendix \ref{Appendix_PLTs_SILCC} and the comment therein). On the other hand, the \rhopdf~of \Htwo~is close to lognormal and no PLT is detected. Examples of the pdfs at two later evolutionary stages are shown in Fig. \ref{fig_rho-pdf_evolution_SILCC} (left). PLTs of the molecular gas well distinguishable from lognormal wings, with slopes $-3\gtrsim q\gtrsim -4$, are detected at $t\sim 13$~Myr (top). At that time both \Hi~and \Htwo~contribute to the PLT of the total gas as read from the derived PLT parameters (cf. Fig. \ref{fig_rho-pdf_evolution_SILCC},  right). 

In the course of further evolution, the molecular gas dominates the density regime at $\rho\gtrsim 10^{-20}$~\gcc~(Fig. \ref{fig_rho-pdf_evolution_SILCC} , bottom left). The latter corresponds mostly to the filamentary clouds, visible in Fig. \ref{fig_SILCC_density_maps}.  The slope of the \rhopdf s, both of \Htwo~and of the total gas, gets shallower and tends to stabilize about $q\gtrsim -1.5$ (bottom right). Its variations around this value are probably due to the complex interplay between gravity, turbulence and external pressure at the late evolutionary stage of the clouds and the formation of prestellar cores therein. We note that the variations of the average slope are more pronounced in the case with \Htwo~due to the smaller density span of the PLT. This behaviour is a characteristic of the adapted \bPLFIT~method -- the lower the DP value is (the larger the PLT span) the less  the slope depends on local features (wiggles etc.) of the PLT. There is a hint of a PLT also for the atomic gas. If real, it is much steeper ($q\sim -3$) than for the molecular gas, but with a similar DP.   

\begin{figure} 
\begin{center}
\includegraphics[width=84mm]{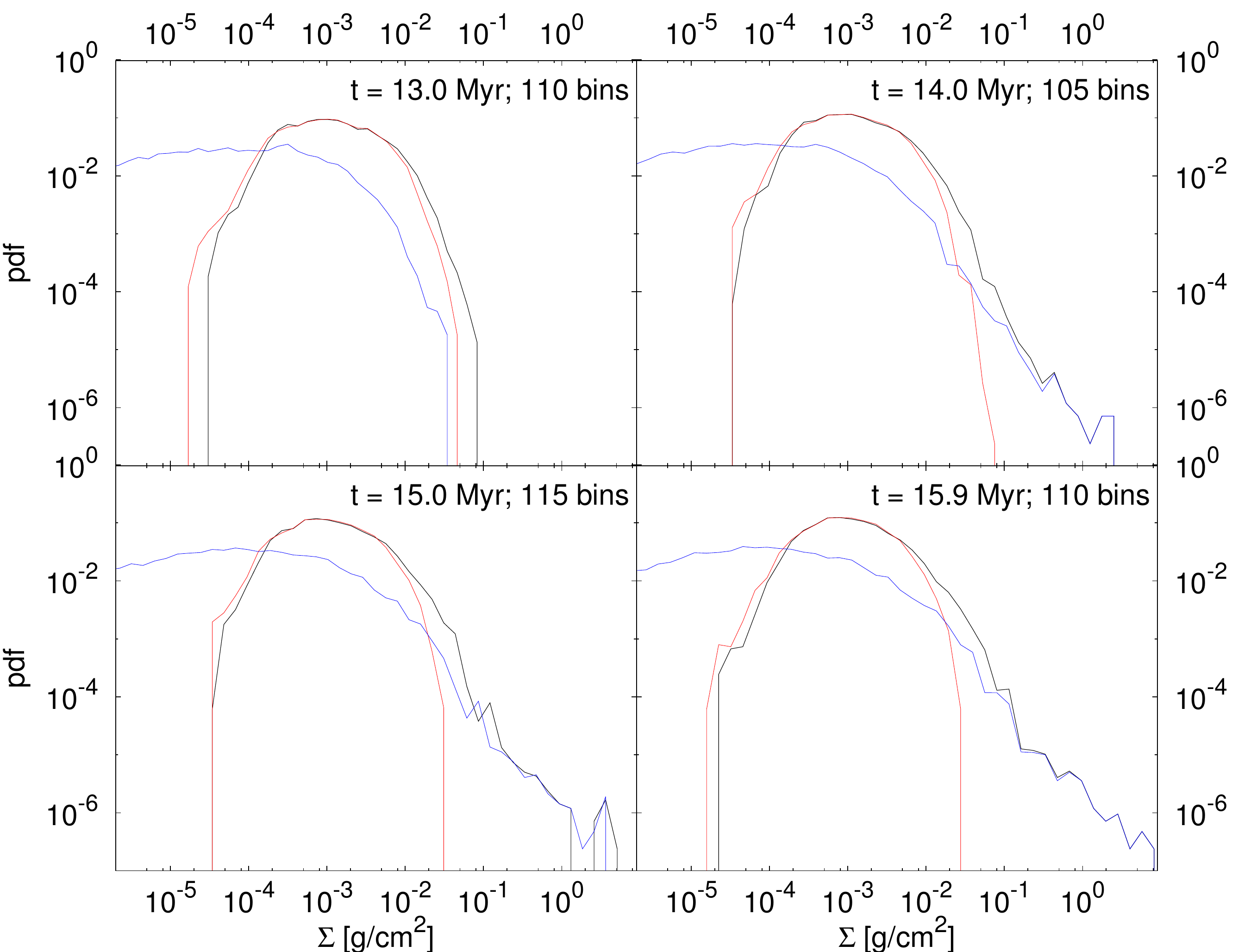}
\caption{Evolution of the \Npdf~as derived from the SILCC simulation. Colors denote the pdf of the total gas (black), of \Hi~(red) and of \Htwo~(blue). }
\label{fig_evolution_N_pdf_examples}
\end{center}
\end{figure}

\begin{figure}
\begin{center}
\includegraphics[width=84mm]{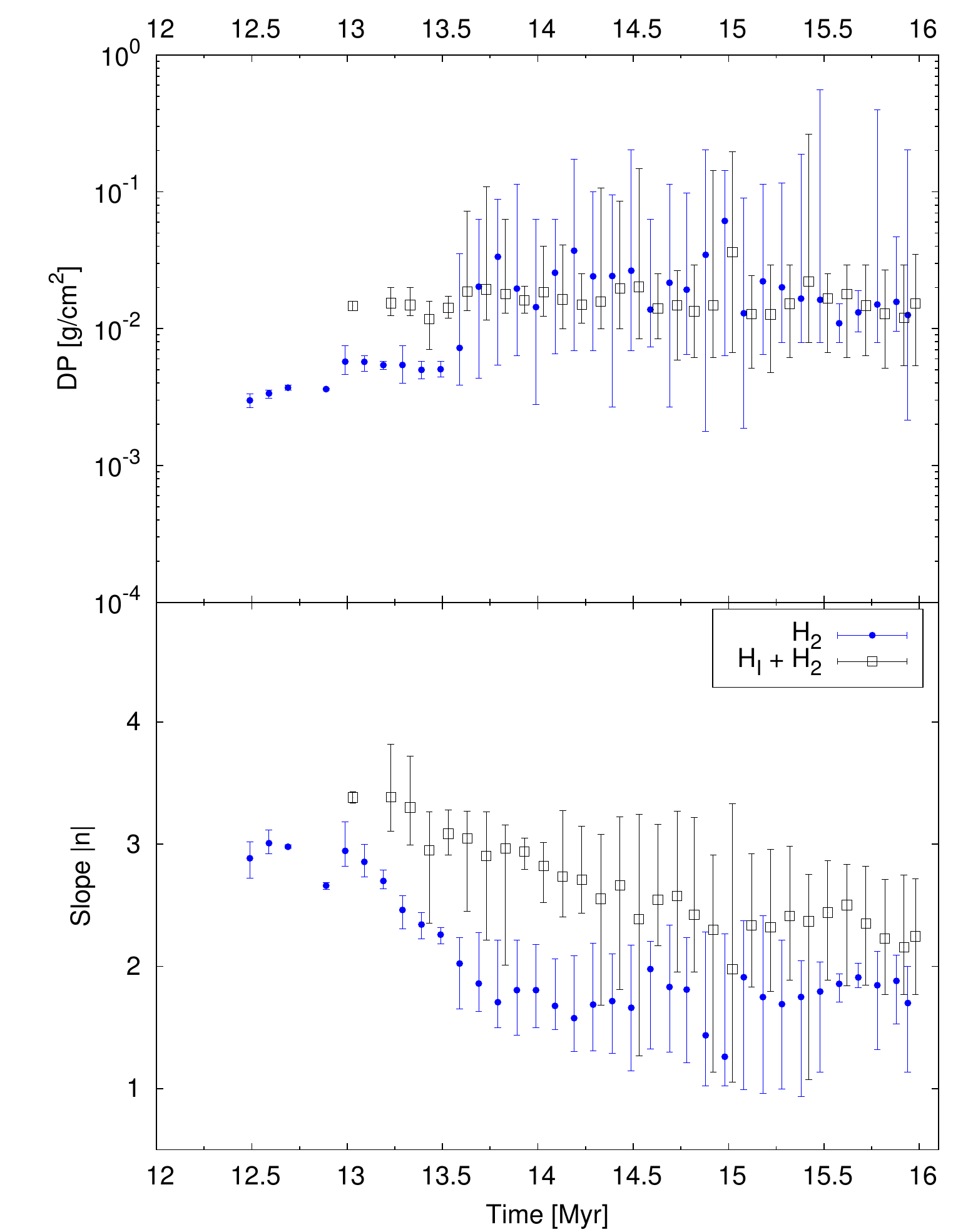}
\caption{Evolution of the \Npdf~PLT parameters from the SILCC data.}
\label{fig_N-pdf_evolution_SILCC}
\end{center}
\end{figure}

We also study the evolution of the \Npdf{} derived from column-density maps along the axis perpendicular to the galactic plane. Fig. \ref{fig_evolution_N_pdf_examples} displays the \Npdf s of the total gas and its atomic and molecular phase at chosen evolutionary times. In contrast to the \rhopdf, the \Npdf~PLT span is almost entirely dominated by molecular gas. The \Npdf~of the atomic gas remains lognormal during the whole studied period, while a PLT in the \Htwo~\Npdf{} emerges as early as $t\lesssim 13$~Myr. The latter develops further as the slope gets shallower and the span extends to over two orders of magnitude. The small spikes which at the high-density end of the PLT are resolution effects and do not influence the slope substantially. The evolution of the \Npdf~parameters for \Htwo~and for the total gas are shown in details in Fig. \ref{fig_N-pdf_evolution_SILCC}. Initially, the slope is steep: $n\sim -3$ for \Htwo; $n\lesssim -3$ for the total gas. At late evolutionary times it gets shallower and seems to approach constant values: around $-2$ for \Htwo, and slightly lower, for the total gas. The DP shifts towards higher densities in the course of evolution of the \Htwo~pdf and varies around a constant value, similar to the stabilization of the slope. On the other hand, the DP of the total-gas \Npdf~is remarkably constant with the evolution. This results from the practically unchanging central, quasi-lognormal part of the atomic-gas \Npdf.

\subsection{Pdf evolution in self-gravitating clumps}
\label{Clump-scale_pdf_evolution}

\begin{figure*} 
\begin{minipage}{\textwidth}
\includegraphics[width=1.\textwidth]{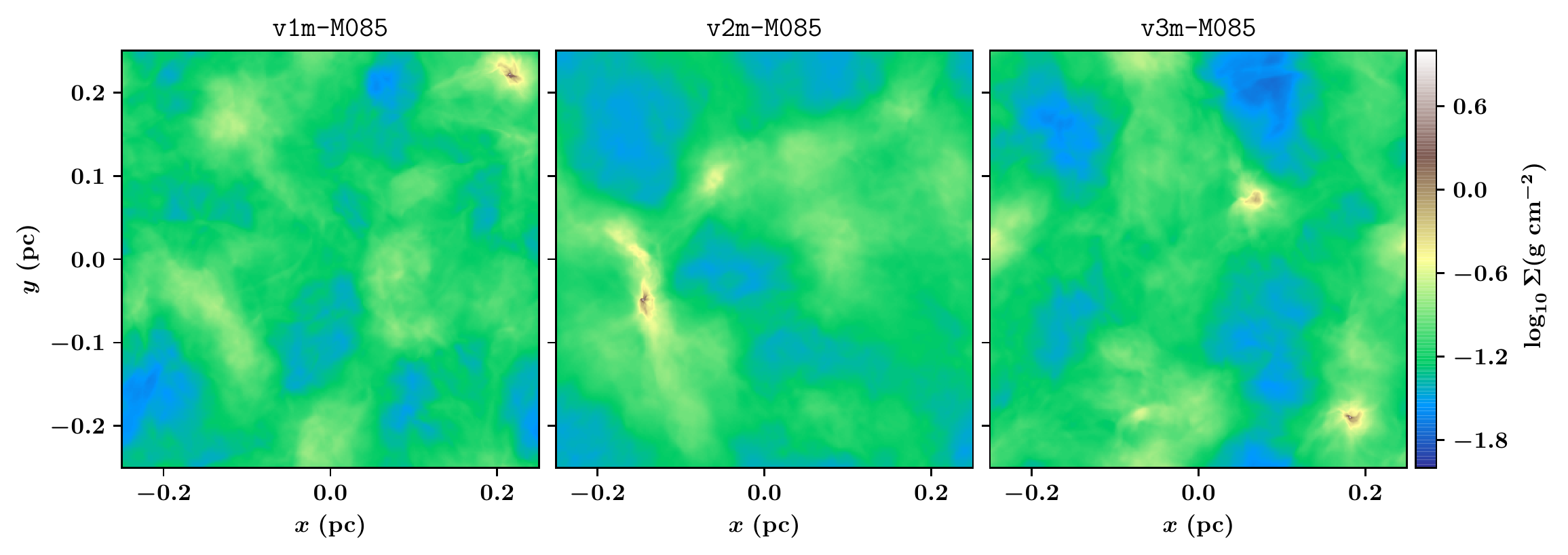}
\includegraphics[width=1.\textwidth]{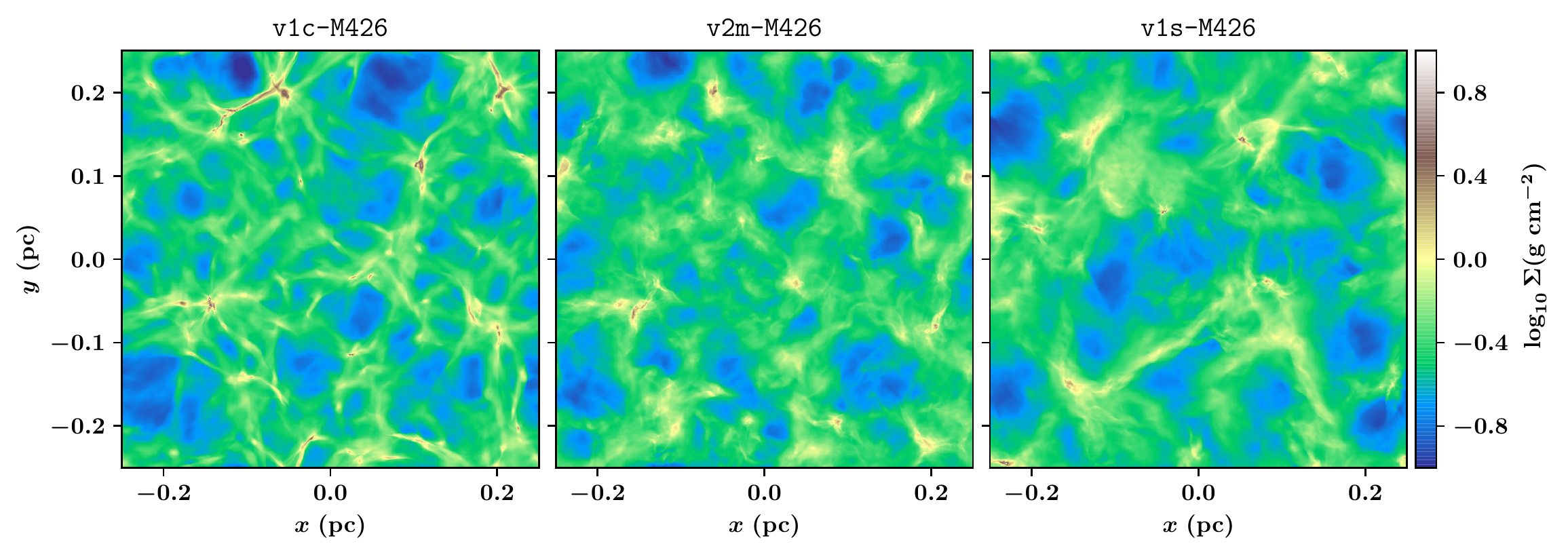}
\caption{Snapshots of the column density at the end of each HRIGT run. The upper panel shows the boxes with lower masses and mixed turbulent driving. In all runs we note the formation of one to a few dense clumps. The bottom panel depicts the runs with higher mass. Here we find numerous clumps, which we attribute mainly to the larger number of Jeans masses in the box. The difference in turbulent driving mode is visible in the filamentary structure. Compressive modes (lower left-hand panel) lead to stronger filaments with larger density contrast. Solenoidal driving (lower right-hand panel) forms more diffuse over-densities. The differences in turbulent driving also lead to significant differences in the total simulation time, see also \citet{GirichidisEtAl2011}. Note the different scales in the column density range.}
\label{fig_HRIGT_density_maps}
\end{minipage}
\end{figure*}

\begin{figure*}
\begin{center}
\includegraphics[width=1.\textwidth]{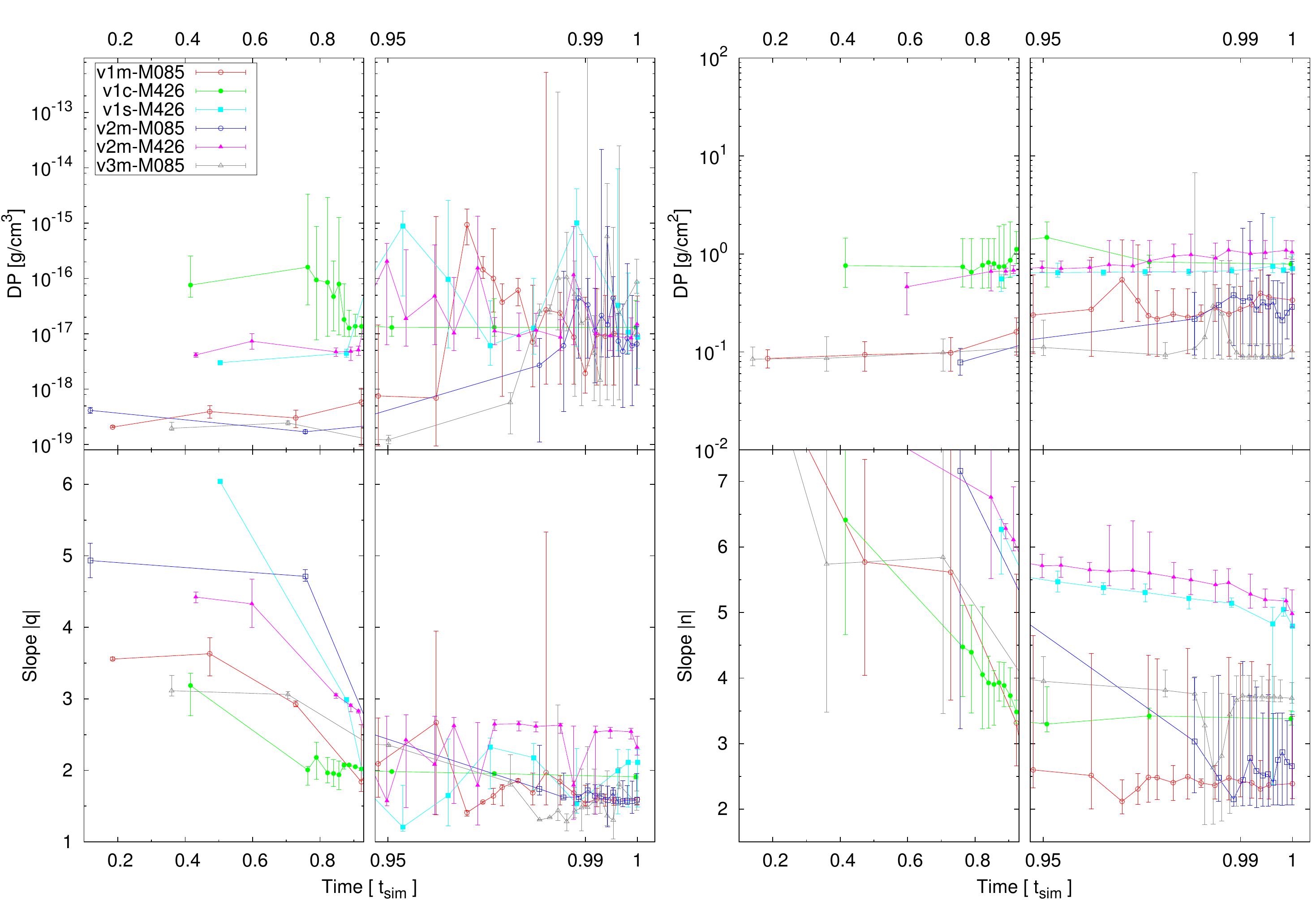}
\caption{Evolution of the parameters of PLTs extracted from the \rhopdf s (left) and the \Npdf s (right) from the HRIGT runs. The time is normalised to the run duration, specified in Table \ref{table_HRIGT_runs}. The time axis for $t\ge 0.9 \langle t\rangle_{\rm ff}$ is stretched exponentially. }
\label{fig_Evolution_pdf_tff_HRIGT}
\end{center}
\end{figure*}

The HRIGT simulations allow us to follow the evolution of \rhopdf~and \Npdf~in self-gravitating media at scales $\le 0.5$~pc, typical for clumps within molecular clouds. Figure~\ref{fig_HRIGT_density_maps} presents the column density maps at the end of the HRIGT runs. Clumps with moderate column density form in the runs with lower density (top panels), with total mass in the box $32~M_{{\rm J},\,0}$  (cf. Table \ref{table_HRIGT_runs}). Filamentary structures containing a number of dense cores appear in the runs with many Jeans masses ($354~M_{{\rm J},\,0}$; bottom panels). The runs with lower total number of Jeans masses (top) show the formation of one to a few overdensities. The majority of the volume is filled with diffuse gas. If the box contains numerous Jeans masses of gas (bottom) relatively weak turbulent velocities are needed to trigger the formation of independently collapsing clumps. The driving mode of the turbulent velocity field results in different times to trigger the collapse. Compressive motions (left-hand panel) push the gas to form filaments with strong density contrast. In contrast, solenoidal forcing (right-hand panel) leads to collapsing clumps embedded in diffuse gaseous structures, see also \citet{GirichidisEtAl2011}. Despite the differences in the spatial gas distribution, the evolution of the PLT parameters in all runs turns out to be quite uniform. It is illustrated in Fig. \ref{fig_Evolution_pdf_tff_HRIGT}. In each run, the initial velocity field introduced in the data cube leads to formation of an approximately lognormal \rhopdf~(not shown) within a few crossing times. A distinguishable PLT with slope $q\gtrsim -4$ appears later in the evolution. At the final stage, as first stars/sink particles form, $q$ seems to vary around some constant value: $\sim -2$ in the runs v1c-M426 and v1s-M426 and $\sim-1.5$, in the other cases (Fig. \ref{fig_Evolution_pdf_tff_HRIGT}, bottom left). On the other hand, the DPs also tend towards some constant value, though with variations up to about one order of magnitude (top left). The latter could be due mostly to foreground and background contamination.

The evolution of the PLT parameters of the \Npdf s is qualitatively similar to the one of the \rhopdf s (Fig. \ref{fig_Evolution_pdf_tff_HRIGT}, right panels). However, the relative duration of the stage characterised by distinguishable PLTs ($n\gtrsim -4$) -- if detected at all -- is much shorter. At this stage the slope shallows rapidly, with a noticeable trend towards some constant value ($n\sim-2.5$) in the runs v1m-M085 and v2m-M085. The range of DP variations at the final evolutionary stage is tighter compared to the \rhopdf{} analysis: about half an order of magnitude in a chosen run.     

\section{Application of the method to {\it Herschel} data}
\label{Results: application to Herschel data}

The adapted \bPLFIT~method was also applied to the \Npdf s in the star-forming regions Aquila and Rosette (Sec. \ref{Herschel data}). Aquila is a filamentary cloud complex of recent star-forming activity where hundreds of starless cores and young protostellar objects have been identified \citep{Konyves_ea_10}. Star formation in the Rosette molecular cloud probably follows two distinct mechanisms. It can be triggered by the expanding \Hii region in a narrow area in the \Hii region/molecular cloud interface zone, but it also takes place deep inside the molecular cloud where filaments merge and provide mass input for the formation of OB clusters \citep{Schneider_ea_12}. The cloud structure is overall highly filamentary and the clumps identified from molecular-line
and dust emission turn out to be mostly gravitationally bound \citep{Veltchev_ea_18}. We can thus expect that the combination of large-scale self-gravity (filament merging) and small-scale gravitational collapse leads to a clear PL tail; see Sec. \ref{Discussion} and the references therein.

\begin{figure} 
\begin{center}
\includegraphics[width=84mm]{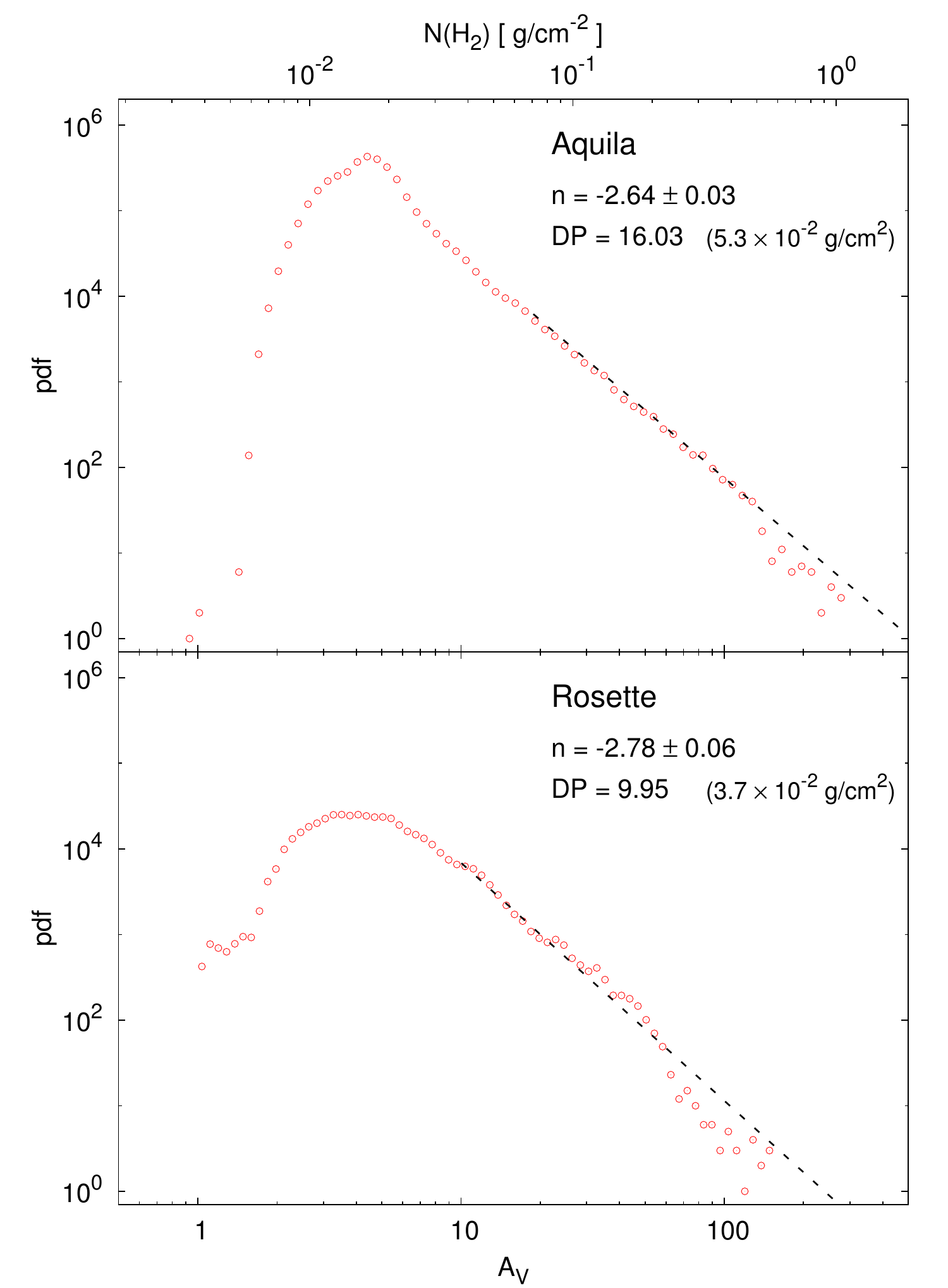}
\caption{Extracted PL tails (dashed lines) from \Npdf s in two Galactic star-forming regions. The conversion to column density is done using the standard factor $N({\rm H}_2)/A_{\rm V} = 0.94 \times 10^{21}$~cm$^{-2}$ mag$^{-1}$ \citep{BSD_78}; mean particle mass of $1.37$ atomic mass units is adopted to calculate the DPs in [ g/cm$^2$ ].}
\label{fig_Observational_N-pdfs}
\end{center}
\end{figure}

Indeed, the PLTs extracted through the adapted \bPLFIT{} from the \Npdf s in Aquila and Rosette are evident and span about one order of magnitude (Fig. \ref{fig_Observational_N-pdfs}). The slopes and the DPs are in general agreement with the values obtained from the SILCC data, for evolved giant molecular clouds (cf. Fig. \ref{fig_N-pdf_evolution_SILCC}). Similar slopes have been found, by use of other techniques, by \citet{Konyves_ea_15} in Aquila ($n=-2.91$) and by Schneider et al. (in prep.) in Rosette ($n=-2.73$). A steeper slope of $n= -3.23$ was found in Rosette by \citet{Tremblin_ea_14} using a lower angular resolution map (36$''$) and fitting a narrower range. We note that the PLT parameters are not affected by local features in the tail. There are some resolution effects at its high-density end but they do not influence the slope, as commented in the next Section. The non-PL part of the \Npdf{} in Rosette has a quasi-lognormal shape while the case in Aquila seems to be more complex. How the output of the adapted \bPLFIT{} compares with PL tails of observational \Npdf s, extracted by use of other approaches (typically, by assuming a combination of lognormal part and a PLT), is worth of a study on its own. We are going to present such comparison in a forthcoming paper (Schneider et al., in preparation).

\section{Discussion}
\label{Discussion}
An obvious advantage of the proposed method is that it rests on a sole assumption about the \rhopdf/\Npdf: {\it the existence of a PLT}. No additional conjectures about the distribution's shape at lower densities are required. The PLT parameters are derived simultaneously which removes the uncertainty of the DP introduced by its assessment by eye. The procedure of averaging of the PLT parameters minimizes the uncertainties caused by some arbitrariness of the binning scheme. Moreover, the adapted \bPLFIT{} method is not sensitive to the low-density completeness limit of the data, as long the latter is lower than the \Npdf{} peak. Therefore the assessed PLT parameters can be considered as reliable. 

A caveat of the method is the unpredictable but -- typically -- rare extraction of false, short PLTs by \bPLFIT{} for some choices of the total number of bins. In fact, this results from fitting at the very {\it high-density} end of the pdf, where the statistic drops. Setting of appropriate maximal span of false PLTs ${\tt FALSE}$ to exclude them solves the problem. Then a reliable assessment of the PLT parameters requires that the tail spans at least one order of magnitude, with a statistically significant number of measurement points providing very low Poisson noise. Poor resolution in dense zones of star-forming regions could lead to distortions or artificial shortening of the tail and hence impede the procedure of averaging of the PLT parameters, even if one sets $(b_{k+1}/b_{\rm min})_{\rm lim}$ as low as a few bins (cf. Section \ref{Results: Adapted method}). 

In case the observed \Npdf s display two PLTs (e.g. \citealt{Russeil_ea_13, Schneider_ea_15b, Pokhrel_ea_16},) the adapted \bPLFIT{} method extracts the DP of the {\it lower-density} one while the slope estimate is not significantly affected by the existence of another PLT. In other words, the method can not extract two PLTs simultaneously -- given the first PLT is detected, the procedure should be reapplied only to the PL part of the \Npdf{}. The choice of the lower density cut-off is critical here and it needs further investigation, in a follow-up work of this study.

The PLT parameters obtained by applying adapted \bPLFIT{} to the selected HRIGT and SILCC data are consistent with the results from other numerical and observational studies. Values of the \rhopdf~slope $q\sim -1.5$, both on galactic and clump scales, are derived at late evolutionary stages of self-gravitating star-forming clouds \citep{Kritsuk_Norman_Wagner_11, Collins_ea_12, Federrath_Klessen_13}, as expected also from theoretical considerations \citep{Girichidis_ea_14, Donkov_Stefanov_18}. On the other hand, \Npdf s derived from observations of regions with star-forming activity display pronounced PLTs of slopes $-2 \gtrsim n \gtrsim -4$ \citep{Abreu-Vicente_ea_15, Schneider_ea_13, Schneider_ea_15a, Pokhrel_ea_16}, also in agreement with our results on the PLT evolution from the SILCC runs (Fig. \ref{fig_N-pdf_evolution_SILCC}).

\begin{figure}
\begin{center}
\includegraphics[width=84mm]{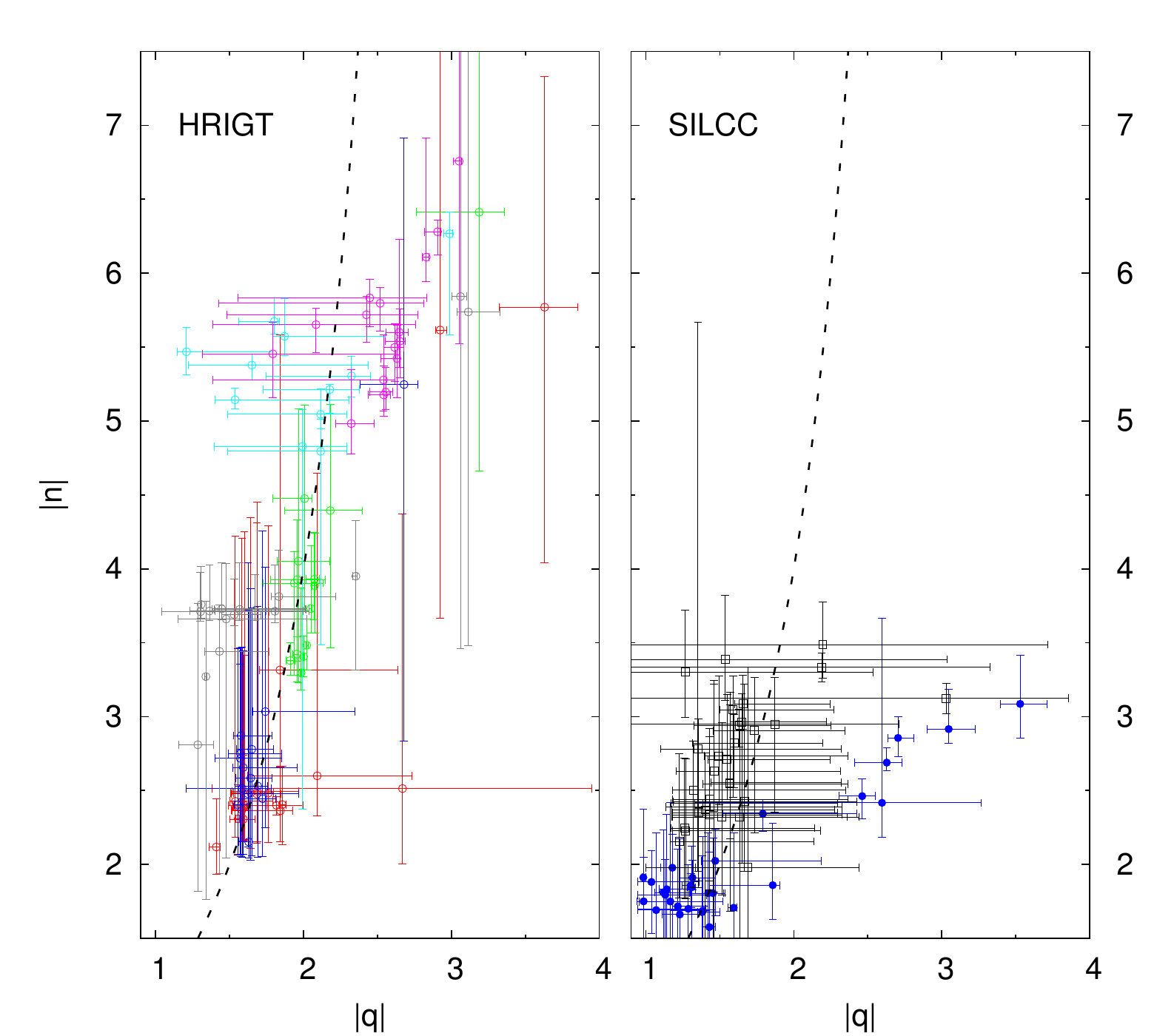}
\caption{Theoretical relation between the PLT slopes of \rhopdf~and \Npdf{} (dashed line; see equation \ref{eq_q-n_relation}), juxtaposed with the results from our numerical runs. The color and symbol notation in the right panel is the same as in Fig. \ref{fig_N-pdf_evolution_SILCC} and in the left one -- like in Fig. \ref{fig_Evolution_pdf_tff_HRIGT}. }
\label{fig_q-n_relation}
\end{center}
\end{figure}

As shown in some of the abovementioned studies, the slope $n$ of the \Npdf~should be related to $q$ as: 
\begin{equation}
\label{eq_q-n_relation}
n=2q/(3+q)~,
\end{equation}
 on the assumption that the general cloud structure can be described through a power-law density profile \citep[see][and the references therein]{Donkov_Veltchev_Klessen_17}. Figure \ref{fig_q-n_relation} presents a test of this relation using the derived values of $q$ and $n$ at different evolutionary times from the HRIGT and SILCC runs. Most of the clump-scale simulations (left) fit well with the prediction of the theory given that the PLTs are clearly distinguishable from lognormal wings, i.e. for slopes $\gtrsim -4$. Note that the latter does not hold for the earliest and relatively short stage characterised by $|q|\gtrsim 2.5$. The slopes derived from the SILCC runs are also in good agreement with formula (\ref{eq_q-n_relation}); again, with a few exceptions at the earliest studied evolutionary stage. A systematic shift of a few dex from the model prediction (dashed line) is evident, mostly considering the pdf evolution of the total gas (Fig. \ref{fig_q-n_relation}, right; black symbols). We interpret this shift in the upward direction -- the \Npdf~PLT slopes $|n|$ are larger than expected while the obtained values of $q$ at later evolutionary times are about the expected one of $-1.5$ \citep{Girichidis_ea_14}. A reasonable explanation might be that at scales about the typical sizes of giant MCs, the relation (\ref{eq_q-n_relation}) changes due to the contribution of collapsing filaments to the PLT of \Npdf. In general, we point out that the adapted \bPLFIT~method extracts PLTs of the \rhopdf s and \Npdf s with slopes which are mutually consistent.     

\section{Summary}
\label{Summary}

We present a new approach to extract the power-law tails (PLTs) of the density (\rhopdf) or column-density distribution (\Npdf) in star-forming clouds. It combines the mathematical method \bPLFIT~\citep{Virkar_Clauset_14} which assesses the power-law part of an arbitrary binned distribution with a procedure of averaging of the output parameters: slope and deviation point (DP), as the number of bins is varied. Reliable results are obtained as long as the PLT spans at least one order of magnitude.

Compared with other techniques for PLT extraction, the proposed adapted \bPLFIT{} method has two considerable advantages. First, it is not based on assumptions about the rest of the pdf (e.g. lognormal) to calculate the slope and the DP -- the sole assumption is that a PLT actually exists. Second, the output parameters are not sensitive to local features (`spikes' etc.) of the distribution tail. 

The adapted \bPLFIT{} was tested on data from numerical simulations of star-forming clouds at clump scale ($0.5$~pc; self-gravitating isothermal medium) and at galactic scales ($500$~pc; stratified gas distribution, chemistry and feedback included). In both cases, the emergence of a PLT takes place at late evolutionary times when protostars (sink particles) and/or dense filamentary clouds are formed. In the course of further cloud evolution, the slopes and the DPs tend to approach constant values -- in agreement with a number of theoretical and numerical studies on the \rhopdf~and \Npdf{} in star-forming regions.

The adapted \bPLFIT{} method was applied also to \Npdf{} of the regions Aquila and Rosette, observed with {\it Herschel}. We found pronounced PLTs with slopes as shallow as $\sim-2.7$, consistent with those for evolved clouds from the used galactic-scale simulation.  

%


\section*{Acknowledgement} We are grateful to the anonymous referee whose critical comments and constructive suggestions helped us to improve this paper.

T.V. and S.D. acknowledge support by the Deutsche Forschungsgemeinschaft (DFG) under grant KL 1358/20-1 and additional funding from the Ministry of Education and Science of the Republic of Bulgaria, National RI Roadmap Project DO1-157/28.08.2018. S.D. acknowledges support by the Bulgarian National Science Fund under Grant N 12/11 (20.12.2017). P.G. acknowledges funding from the European Research Council under ERC-CoG grant CRAGSMAN-646955. T.V. and N.S. acknowledge support from the ANR/DFG collaboration project GENESIS (ANR-16-CE92-0035-01/DFG1591/2-1). O.S. and L.M. acknowledge support by the Scientific Research Fund of the University of Sofia under Grant \#80-10-68/19.04.2018. R.S.K. thanks funding from the DFG in the Collaborative Research Center (SFB 881) "The Milky Way System" (subprojects B1, B2, and B8) and in the Priority Program SPP 1573 "Physics of the Interstellar Medium" (grant numbers KL 1358/18.1, KL 1358/19.2) D.S. acknowledges the support of the Bonn-Cologne Graduate School, which is funded through the German Excellence Initiative. D.S. also acknowledges funding by the DFG via the Collaborative Research Center SFB 956 “Conditions and Impact of Star Formation” (subprojects C5 and C6). 

The {\sc Flash} code used in this work was partly developed by the Flash Center for Computational Science at the University of Chicago. The authors acknowledge the Leibniz-Rechenzentrum Garching for providing computing time on SuperMUC via the project “pr94du” as well as the Gauss Centre for Supercomputing e.V. (www.gauss-centre.eu).

\label{lastpage}
\newpage
\appendix
\section{Column-density maps}
\label{Maps of the SF regions}
\begin{figure} 
\begin{center}
\includegraphics[width=75mm]{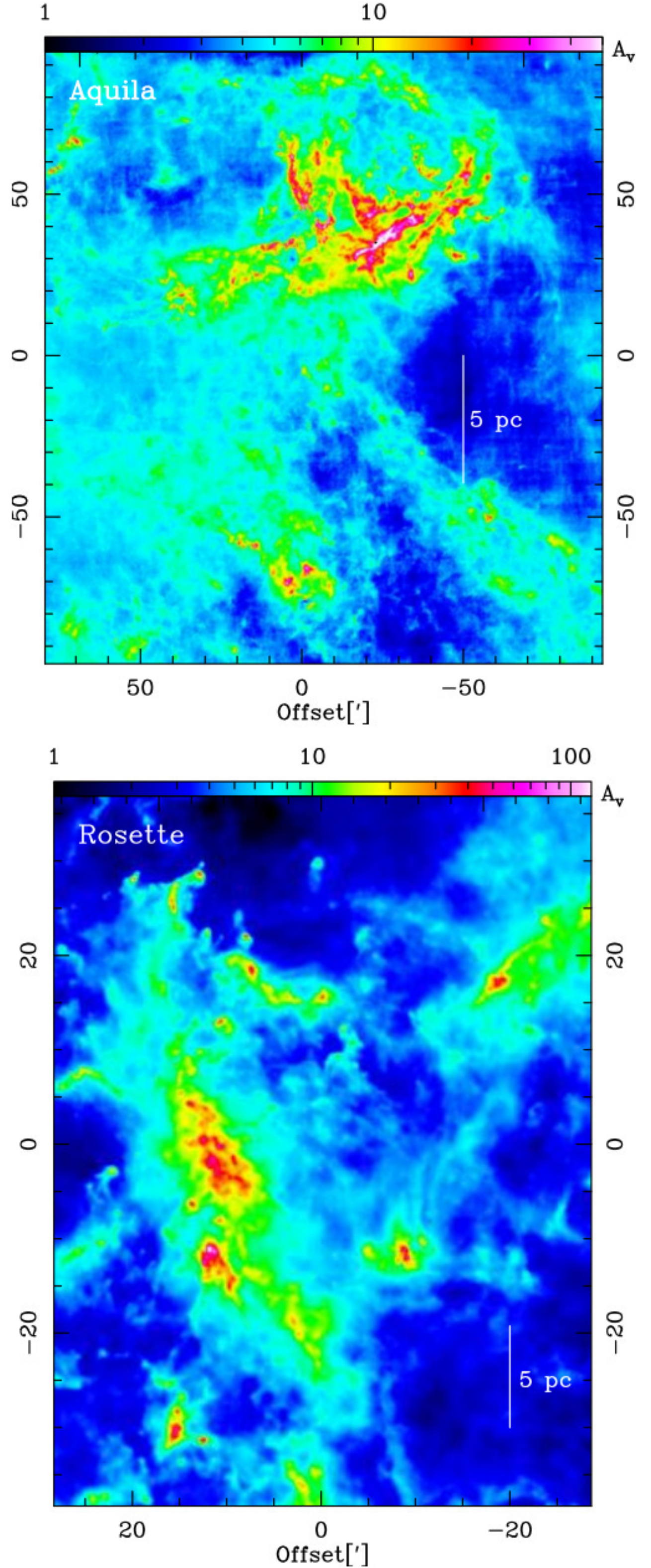}
\caption{{\it Herschel} maps of Aquila (top) and Rosette (bottom) used for construction of the corresponding $N$-pdfs (cf. Section \ref{Herschel data}).}
\label{fig_Herschel_maps}
\end{center}
\end{figure}

Column-density maps of the studied star-forming regions are shown in Fig. \ref{fig_Herschel_maps}. Note that they are rotated so that no absolute coordinates are given. We express the column density in visual extinction, using the conversion $N(H) = A_{V} 0.94 \times 10^{21}$~cm$^{-2}$mag$^{-1}$ \citet{BSD_78}. 

\section{Test of \bPLFIT{} on analytical pdfs}
\label{Appendix_bPLFIT_analytical pdfs}

\begin{figure} 
\begin{center}
\includegraphics[width=84mm]{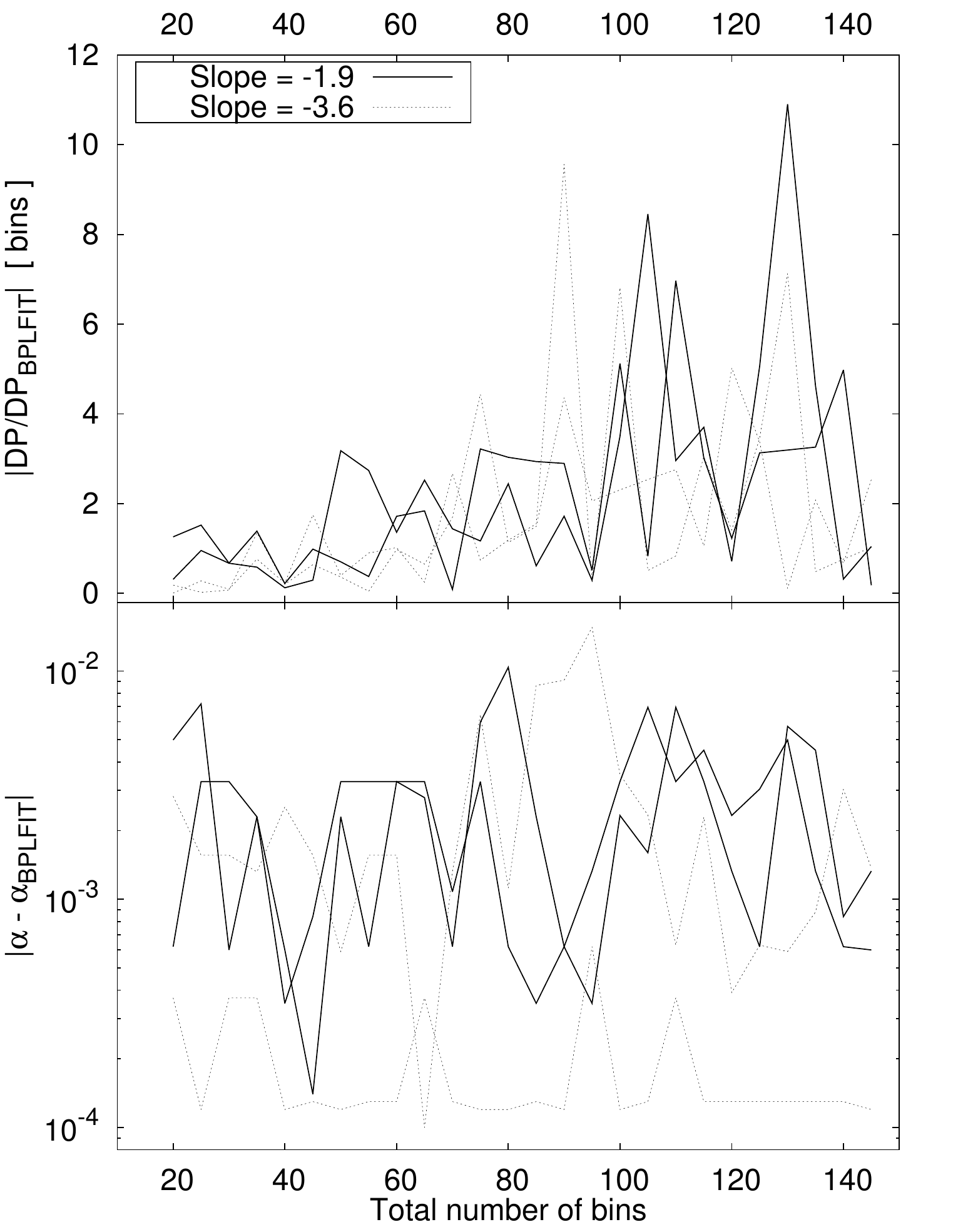}
\caption{Test of \bPLFIT{} on analytical pdfs with known PLT parameters. See text.}
\label{fig_bPLFIT_test}
\end{center}
\end{figure}

We test the accuracy of the \bPLFIT{} technique by applying it to two series of analytical pdfs which are combinations of a lognormal function and a PLT. Pdfs with a PLT slope $-3.6$ (short, unevolved tail) are illustrative for early evolutionary state of the cloud while those with slope $-1.9$ are typical for later evolutionary stages. The results for two pairs from both groups are shown in Fig. \ref{fig_bPLFIT_test}; each pair is characterized by the same slope while the PLT span is different. The deviations of the extracted PLT parameters from the real values do not seem to depend on the total number of bins $k$. The slopes are derived with a high average accuracy of $10^{-3}$ (bottom panel) whereas the typical deviation of the DP from the real value (top panel) is between 1 and 4 bins though it may be larger in some exceptional cases.  

\section{PLTs at an early evolutionary stage}
\label{Appendix_PLTs_SILCC}

\begin{figure} 
\begin{center}
\includegraphics[width=83mm]{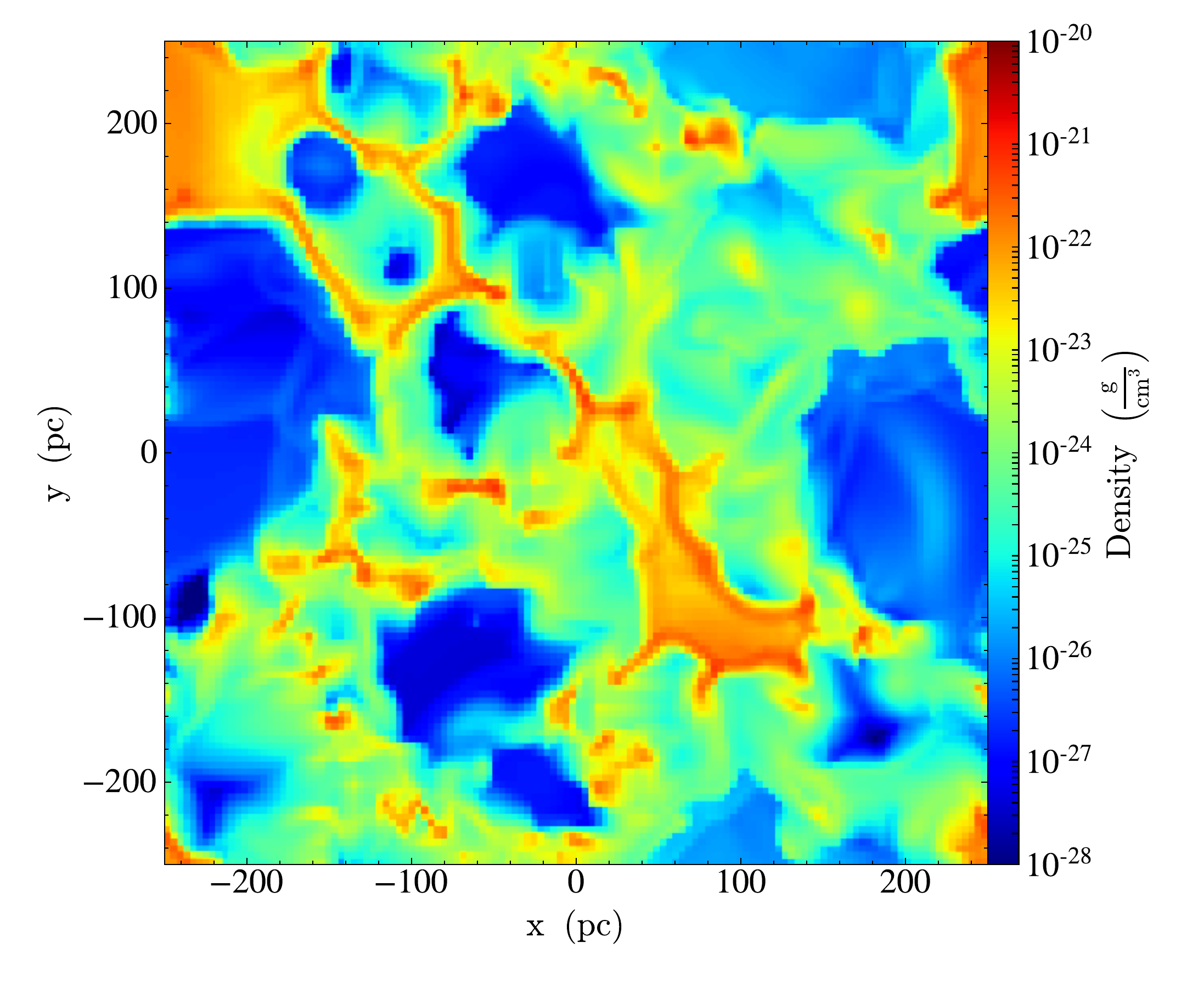}\\
\includegraphics[width=83mm]{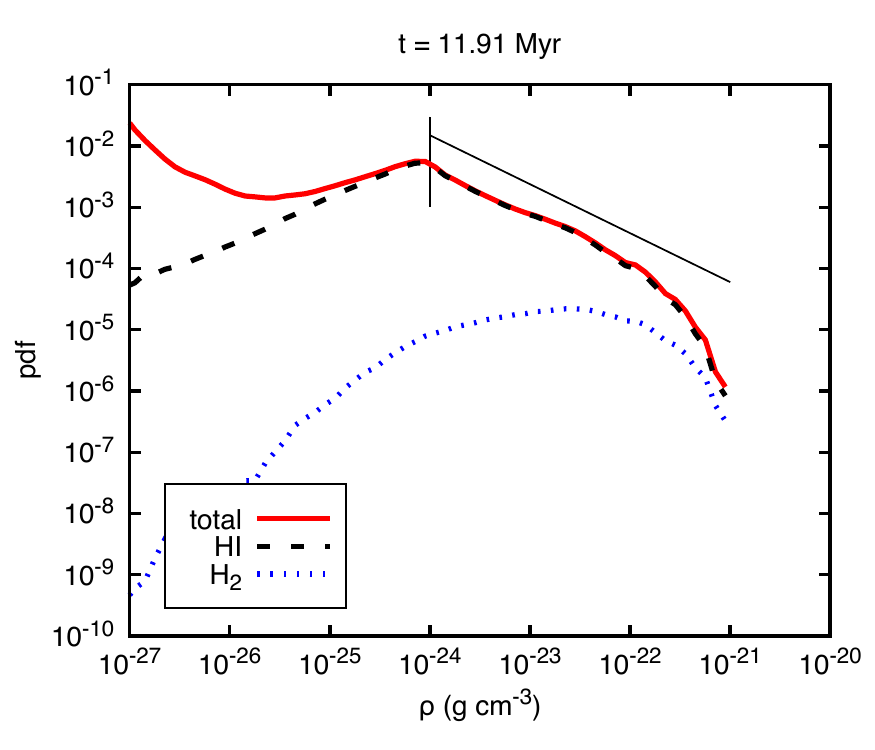}
\caption{Slice of the density field ($z=0$) of the total  gas (\Hi{} + \Htwo + H$^+$) in the SILCC run (top) and the corresponding \rhopdf s of different tracers (bottom) before we start the additional refinements in the zoom-in regions. The DP (short vertical mark) and the shallow slope (dashed line) are denoted. See text.}
\label{fig_PLT_early_evolution}
\end{center}
\end{figure}

As seen in Fig. \ref{fig_rho-pdf_evolution_SILCC}, the \rhopdf s of the atomic and the total gas at early stages of the SILCC run are characterized by very shallow PLTs ($q>-1$). An example is shown in Fig. \ref{fig_PLT_early_evolution}, bottom. The DP coincides with a pronounced maximum of the distribution and corresponds to a typical density of the diffuse zones ($\rho\sim 10^{-24}$~\gcc) which separates large `voids' of very diluted, mainly atomic gas from the emerging giant MCs (cf. top panel). The \rhopdf{} at this stage is still influenced by the initial conditions. The initial setup results in a flat density \rhopdf{} with an initial peak at $\sim 10^{-23}\,\mathrm{g\,cm^{-3}}$. During the initial phase with cooling and a vertical contraction of the disc a non-negligible fraction of the gas is compressed to higher densities. At the same time the SN driving creates over-densities via shock waves, which results in an accumulation of gas at $\sim 10^{-24}\,\mathrm{g\,cm^{-3}}$. The emerging peak in the pdf at that density effectively leads to flat slopes for the higher densities. As this behaviour mainly reflects the initial conditions rather than any physical processes in equilibrium or steady state we should not overinterpret the value of the slope.

%
\end{document}
\fi